\documentclass[preprint,showpacs,preprintnumbers,amsmath,amssymb]{revtex4}
\usepackage{graphicx}
\usepackage{dcolumn}
\usepackage{bm}
 
\begin{document}
\title{Intriguing electron correlation effects in the photoionization of metallic 
quantum--dot nanorings}

\author{Ioan B\^aldea}
\email{ioan@pci.uni-heidelberg.de}
\altaffiliation[Also at ]{National Institute for Lasers, 
Plasmas, and Radiation Physics, ISS,
RO--76900 Bucharest--M\u{a}gurele, Romania.}
\author{Lorenz S.\ Cederbaum}
\author{Jochen Schirmer}
\affiliation{
Theoretische Chemie,
Physikalisch--Chemisches Institut, Universit\"{a}t Heidelberg, Im
Neuenheimer Feld 229, D--69120 Heidelberg, Germany}

\date{\today}

\begin{abstract}
We report detailed results on ionization in metallic quantum--dot (QD) nanorings 
described by the extended Hubbard model at half filling 
obtained by exact numerical diagonalization.
In spite of very strong electron correlations, the ionization spectra  
are astonishingly scarce. We attribute this scarcity to a hidden quasi--symmetry, 
generalizing thereby 
similar results on optical absorption recently reported [I.\ B\^aldea and L.\ S.\ Cederbaum, 
Phys.\ Rev.\ B {\bf 75}, 125323 (2007); {\bf 77}, 165339 (2008)].
Numerical results indicate that this hidden quasi--symmetry of the extended Hubbard model 
does not evolve into a true (hidden) symmetry but remains a quasi--symmetry in the case of 
the restricted Hubbard model as well.
Based on the observation on the number of 
significant ionization signals per each spatial symmetry, 
we claim the existence of a one--to--one map 
between the relevant ionization signals of the correlated half-filled nanorings 
and the one-hole and two-hole--one-particle processes possible in the noninteracting case.
Similar to the case of optical absorption, numerous avoided crossings (anticrossings) 
are present in the ionization spectra, which often involve more than two states.
The present results demonstrate that ionization could be a 
useful tool to study electron correlations in metallic QD--nanoarrays, providing information that 
is complementary to optical absorption.  
\end{abstract}
 

\maketitle

\renewcommand{\topfraction}{1}
\renewcommand{\bottomfraction}{1}
\renewcommand{\textfraction}{0}
\section{Introduction}
\label{sec:introd}
Metallic quantum dots (QDs) can be fabricated and assembled in extended 
regular arrays by means of modern nanotechnologies 
\cite{Heath:97a,Collier:97,Ingram:97,Markovich:98,Shiang:98,Chen:98,Medeiros:99}. 
The salient feature of such artificial nanostructures is the fact that 
their electronic properties can be 
smoothly \emph{tuned} in wide ranges by factors that can be easily controlled experimentally:
dot diameter $2R$, interdot spacing $D$, and gate potentials. 
It is this tunability of nanostructures that makes the most important difference from 
ordinary molecules and solids.
In isolated QDs with sizes of a few nanometers, typical for the nanograins of silver prepared by 
Heath' 
group \cite{Collier:97,Heath:97a,Markovich:98,Shiang:98,Medeiros:99,Henrichs:00,Sampaio:01,Beverly:02}.
the single--electron levels, characteristic for quantum boxes or ``artificial'' atoms 
\cite{Kastner:93}, 
are well separated by energies $\sim 1$\,eV. 
A small fraction of ``valence'' electrons, which occupy the highest ``atomic'' orbitals, 
becomes delocalized over the whole nanostructure, if the QDs are sufficiently densely packed  
and the adjacent electronic wave functions overlap.  

For nearly touching Ag--QDs ($d\equiv D/2R \agt 1$), correlations of these delocalized electrons 
are negligible: the single--particle, molecular--orbital (MO) picture represents a very good 
approximation \cite{Baldea:2002,Baldea:2007,Baldea:2008}. This description progressively worsens 
for more and more distant QDs, and completely breaks down 
towards the end of the $d$-range of experimental interest ($d \alt 2$).
Electron correlations become more and more important beyond 
$d \agt 1.4$ \cite{Baldea:2002,Baldea:2004a,Baldea:2007,Baldea:2008}.
The attractive fact to study such QD--nanostructures is that by varying $d$ 
one can drive \emph{continuously} a many--electron system 
between weak and strong correlation regimes.
Obviously, this route from weak to strong correlation regimes is inconceivable in 
ordinary molecules and solids.

The present paper is devoted to metallic QD nanorings. 
To our knowledge, rings of metallic QDs have not 
yet been assembled so far. For this purpose, the use of a 
mask in the preparation of a two--dimensional nanoarrays 
\cite{Collier:97,Heath:97a,Markovich:98,Shiang:98,Medeiros:99,Henrichs:00,Sampaio:01,Beverly:02}
is conceivable.
As shown recently \cite{Baldea:2007,Baldea:2008},
electron correlations in these metallic QD--nanorings turned out to possess an 
intriguing character. In spite of very strong correlations, the optical 
absorption spectra are astonishingly scare (almost monochromatic for closed shells \cite{Baldea:2007}) 
and can be rationalized within a single--particle description. It would be 
interesting to further investigate to what extent the hidden quasi--symmetry 
suggested in the context of optical absorption \cite{Baldea:2007,Baldea:2008}
manifests itself in other properties. Along this line, in the present investigation 
we shall focus our attention to (photo)ionization, a property 
for which electron correlations are known to be important 
\cite{Siegbahn:1967,Siegbahn:1969,Siegbahn:1982,Turner:1970,Rabalais:1977,Berkowitz:1979,Cederbaum:1977b,Cederbaum:1980,Cederbaum:1986}.
The important role played by electron correlation in the ionization of QD nanorings became evident 
in the preliminary study of Ref.\ \onlinecite{Baldea:2002}. In this sense, the present work in an 
extension of the investigation of Ref.\ \onlinecite{Baldea:2002}, 
where only the lowest energy process was considered. This is the so--called HOMO process,
because within the single--particle picture it would correspond to removing an electron from the 
{\bf h}ighest {\bf o}ccupied {\bf m}olecular {\bf o}rbital. 
The fact demonstrated in Ref.\ \onlinecite{Baldea:2002}, 
that the molecular orbital picture of ionization completely breaks down in QD--nanorings even 
for the HOMO--processes is very interesting, because it is contrary to ordinary molecules, where it 
holds; see Refs.\ \onlinecite{Siegbahn:1967,Siegbahn:1969,Siegbahn:1982,Turner:1970,Rabalais:1977,Berkowitz:1979}. 
This represents a further motivation for studying ionization in QD--nanorings.
\par
To anticipate, the results for ionization we are going to present below fully support
the existence of a hidden quasi--symmetry in the strongly correlated 
metallic QD-nanorings described 
within the extended Hubbard model found in optical absorption \cite{Baldea:2007,Baldea:2008}.
Still, because we cannot present an analytical proof on the existence of a 
hidden quasi-symmetry, we believe that it is important and beneficial for further investigations
to amply document it by detailed exact results  
obtained by means of extended numerical calculations.
Because the numerical study is rather intricate, in Sec.\ \ref{sec:model},
where the model of QD--nanorings is exposed, we also provide significant details,
which were not given in the earlier studies. 
In the next Sec.\ \ref{sec:correlations} we analyze the electron correlations 
from a standpoint different from that adopted in Refs.\ 
\cite{Baldea:2002,Baldea:2007,Baldea:2008}, \emph{e.~g.},~by considering 
the quantum entanglement. 
In Sec.\ \ref{sec:results}, the results on ionization spectra 
will be presented. 
We shall examine in detail the cases of nanorings consisting of six 
and ten QDs separately, in Secs.\ \ref{sec:6-QDs} and \ref{sec:10-QDs}, 
respectively. 
The two main effects of electron correlations, the avoided crossings and the scarcity 
of the ionization spectra will be considered in Secs.\ \ref{sec:avoided-crossings} 
and \ref{sec:hidden-symmetry}, respectively.  
Conclusions will be presented in the final Sec.\ \ref{sec:conclusion}.
\section{Model and method}
\label{sec:model}
To describe the valence electrons of metallic QD--nanorings, 
we shall utilize, similar to a series of previous studies
\cite{Remacle:2000a,Baldea:2002,Baldea:2004a,Baldea:2007,Baldea:2008},
an extended Hubbard (or Pariser--Parr--Pople) model Hamiltonian
\begin{eqnarray}  
H = & - & t_{0} \sum_{l=1}^{N}    
\sum_{\sigma=\uparrow,\downarrow}  
\left(a_{l,\sigma}^{\dagger} a_{l+1,\sigma}^{} +   
a_{l+1,\sigma}^{\dagger} a_{l,\sigma}^{}\right) \nonumber \\  
{} & + &  
\sum_{l=1}^{N} \left( \varepsilon_H \, n_{l}^{ } + 
U n_{l,\uparrow}^{} n_{l,\downarrow}^{} +  
V n_{l}^{} n_{l+1}^{}\right),
\label{eq-model-hamiltonian}  
\end{eqnarray}  
where, $a$ ($a^{\dagger}$) denote creation   
(annihilation) operators for electrons,  
$n_{l,\sigma} \equiv a_{l,\sigma}^{\dagger} a_{l,\sigma}^{}$,  
$n_{l} \equiv n_{l,\uparrow} + n_{l,\downarrow}$,
The ideal situation assumed in Eq.\ (\ref{eq-model-hamiltonian}), 
where the model parameters are site independent, can be 
considered a reasonable approximation in view 
of the narrow size distributions ($\sim 2 - 5\%$) in the arrays of Ag QDs 
assembled by Heath's group 
\cite{Heath:97a,Collier:97,Markovich:98,Shiang:98,Medeiros:99}.
As previously discussed \cite{Baldea:2002,Baldea:2007}, 
such a weak disorder does not significantly alter the results obtained 
by assuming site--independent parameters.
Importantly for subsequent considerations, the point symmetry group associated with 
the model (\ref{eq-model-hamiltonian}) of nanoring is D$_{Nh}$.
\par
The model parameter entering Eq.\ (\ref{eq-model-hamiltonian}) 
has been analyzed earlier in literature 
\cite{Collier:97,Medeiros:99,Remacle:98a,Remacle:98b,Remacle:01,Remacle:02a,Baldea:2002}
and will be therefore no more repeated here.
Importantly, the fact that the interdot separation $d\equiv D/(2R)$ 
(measured between QD centers) can be continuously varied 
in the range $1.10 \lesssim d \lesssim 1.85$ by means of a Langmuir technique 
allows a wide parameter tuning 
\cite{Heath:97a,Collier:97,Markovich:98,Shiang:98,Medeiros:99}.
For concrete values, see Fig.\ 1 of Ref.\ \onlinecite{Baldea:2008}.
At the ends of the aforementioned $d$-range,
the ratio of the on--site repulsion $U$ to the free electron 
bandwidth $4 t_{0}$ varies from values $U/(4 t_{0}) \ll 1$ to $U/(4 t_{0}) \gg 1$. 
Therefore, a crossover from a weakly 
correlated system to a strongly correlated one by varying $d$ can be reached.
\par
Ionization is usually described as ejection of an electron from a 
``molecular'' orbital (MO). Therefore, besides electron operators $a_{l,\sigma}$ 
for ``atomic'' orbitals, it is also useful to introduce operators 
of MOs (or Bloch states),
$c_{k,\sigma} \equiv 1/\sqrt{N} \sum_{l} a_{l,\sigma} \exp(-2\pi k l i/N)$.
An ionization signal is characterized by an ionization potential 
$\varepsilon_{k,i} = \langle \Psi_{k, i}\vert H \vert \Psi_{k, i}\rangle - 
\langle \Phi\vert H \vert \Phi\rangle $ and a spectroscopic factor $w$
\begin{equation}
\displaystyle
w_{k, i} = 
\vert\langle \Psi_{k, i}\vert c_{k,\sigma}\vert 
\Phi \rangle\vert^{2}. 
\label{eq-w}
\end{equation} 
Here, $k$ denotes the wave number of the Bloch state (MO) out of which 
the electron is removed and $\Psi_{k, i}$ ($i=1, 2,\ldots$, see below for notation) 
are eigenstates of the ionized 
nanoring. We shall only consider ionization at zero temperature, that is, 
the nonionized system is initially prepared in its ground state $\Phi$. 
The quantities $w_{k,i}$ are important, since the photoionization signals are  
proportional to them \cite{Cederbaum:1977b,Cederbaum:1980,Cederbaum:1986}.
\par
Symmetry considerations play an important part for ionization in nanorings,
similar to optical absorption \cite{Baldea:2007,Baldea:2008}. 
In the latter case, the counterpart of the 
matrix elements in the r.h.s.\ of Eq.\ (\ref{eq-w})
entering the coefficient of optical absorption are the elements of 
the dipole operator $\hat{\mathbf{P}}$.
For the case of closed shells considered here, we have always found that 
the ground state $\Psi$ possesses $A_{1g}$ symmetry. 
In this case, because the dipole operator $\hat{\mathbf{P}}$ possesses $E_{1u}$ 
symmetry \cite{Baldea:2007,Baldea:2008}, only $E_{1u}$--excited eigenstates 
can be targeted in optical absorption. To target eigenstates with symmetries different 
from $E_{1u}$ by zero--temperature optical absorption, one needs to consider 
open shell rings, characterized by ground state symmetries $\Gamma_{\Phi} \neq A_{1g}$, as 
done in Ref.\ \onlinecite{Baldea:2008}. The situation is different for ionization. 
The MOs span many irreducible representations $\Gamma_{k}$ of the point group 
D$_{Nh}$ \cite{MO-symmetries}. Eigenstates 
$\Psi_{k, 1}, \Psi_{k, 2}, \Psi_{k, 3}, \ldots $ of all MO-symmetries (specified by the value of $k$)
can be targeted by ionization even with an initial neutral $A_{1g}$--state $\Phi$
(because $\Gamma_{k} \otimes \Gamma_{k} \supset A_{1g}$). To this, it suffices to consider 
nanorings with closed shells. For the sake of simplicity, we shall restrict ourselves here 
to the prototypical case of closed shells, the half--filling case. 
\par
Numerical results will be presented for half--filled nanorings with six and 
ten QDs. For the former case, all eigenstates of Hamiltonian (\ref{eq-model-hamiltonian})
can be computed exactly by numerical diagonalization: the dimension of the 
total Hilbert space is equal to 924 for the neutral ring and 792 for the ionized ring. 
The neutral ground state is a singlet, and one can carry out calculations in subspaces 
with total spin projection $S_z=0$ and $S_z=+1/2$, respectively. 
The corresponding dimensions are then reduced to 400 and 300, respectively.
The fact that \emph{all} eigenstates of Hamiltonian (\ref{eq-model-hamiltonian})
can be computed exactly enables us a very detailed analysis of all 
relevant physical aspects.
For ten QD--nanorings, the total dimension is equal to 184756 in neutral rings and 
167960 in ionized rings. Similar to the above case, the restriction 
to the subspaces with $S_z=0$ and $S_z=+1/2$ is also possible, which results in the reduced 
dimensions of 63504 for neutral rings and 52920 for ionized rings. For such dimensions, 
the Lanczos algorithm has been applied. By running three 
times the Lanczos procedure we are able to compute 
the ionization energies and the spectroscopic factors individually \cite{koeppel:84}.
The difference from calculations for optical 
absorption \cite{Baldea:97,Baldea:2004b,Baldea:2007,Baldea:2008}  
is that the first two Lanczos runs are carried out for the nonionized ring, 
while in the third run one must choose the normalized vector 
$c_{k,\sigma}\vert\Phi\rangle$ as starting Lanczos vector separately for each symmetry $\Gamma_{k}$
and perform calculations for the ionized ring.
\par
It is worth noting at this point that the continued fraction algorithm, familiar 
in condensed matter physics \cite{HHK:80,fulde:91,dagotto:94}, 
does not suffice for the present purpose, since it only 
allows calculations of convoluted spectra. The phenomenon of avoided crossing, 
often encountered in the study on optical absorption in QD nanorings\cite{Baldea:2008} 
and also important for ionization 
(see Secs.\ \ref{sec:6-QDs}, \ref{sec:10-QDs}, and \ref{sec:avoided-crossings}),
represents a typical situation where the information on convoluted spectra 
obtained by this algorithm is totally insufficient. 
\par
Especially for the case of larger nanorings, for which the Lanczos algorithm 
has to be applied, 
the following sum rule, straightforwardly deduced from Eq.\ (\ref{eq-w}),
\begin{equation}
\displaystyle
\sum_{i} w_{k, i} = 
\langle \Phi \vert c_{k,\sigma}^{\dagger}  c_{k,\sigma} \vert \Phi\rangle \equiv n_{k,\sigma}. 
\label{eq-sum-rule}
\end{equation} 
turns out to be very useful as check of numerical calculations.
\par
At the end of this section, we note that, 
similar to earlier works \cite{Baldea:2007,Baldea:2008}, 
we safely rule out any spurious contribution, \emph{e.~g.}, 
those of higher--spin eigenstates.
Because the ket state $c_{k,\downarrow}\vert \Phi \rangle$
entering Eq.\ (\ref{eq-w}) is an exact eigenstate of both the total spin projection 
and the total spin, $S_z = S = 1/2$ \cite{total-spin}, only the 
spin doublet eigenstates $\Psi_{k, i}$ can possess nonvanishing 
spectroscopic factors $w_{k, i}$. 
For the ionized six--QD nanorings, out of the total 300 eigenstates with $S_z=1/2$ 
we only retained the 210 eigenstates with $S_z=S=1/2$. In the calculations 
for the ionized ten--QD nanorings, the starting Lanczos vector possesses the 
correct spin (see above), but accumulated numerical errors during 
the Lanczos iterations could yield Ritz vectors containing unphysical admixtures
with other eigenstates. We can unambiguously exclude this possibility: 
we have stored \emph{all} the relevant Ritz vectors 
(usually at most 100 vectors of dimensions 52920) 
and checked that they are very accurate 
eigenvectors with correct spin, and exhaust Eq. (3). 
Their dispersion was always found to be very small, comparable to 
that of the neutral eigenstate $\Phi$. 
In particular, it is at least 3 -- 4 orders of magnitude smaller than 
the energy difference at avoiding crossings (see Sec.\ \ref{sec:avoided-crossings}).
\section{Electron correlations}
\label{sec:correlations}
\par
\begin{figure}[h]
\centerline{
\includegraphics[width=0.35\textwidth,angle=-90]{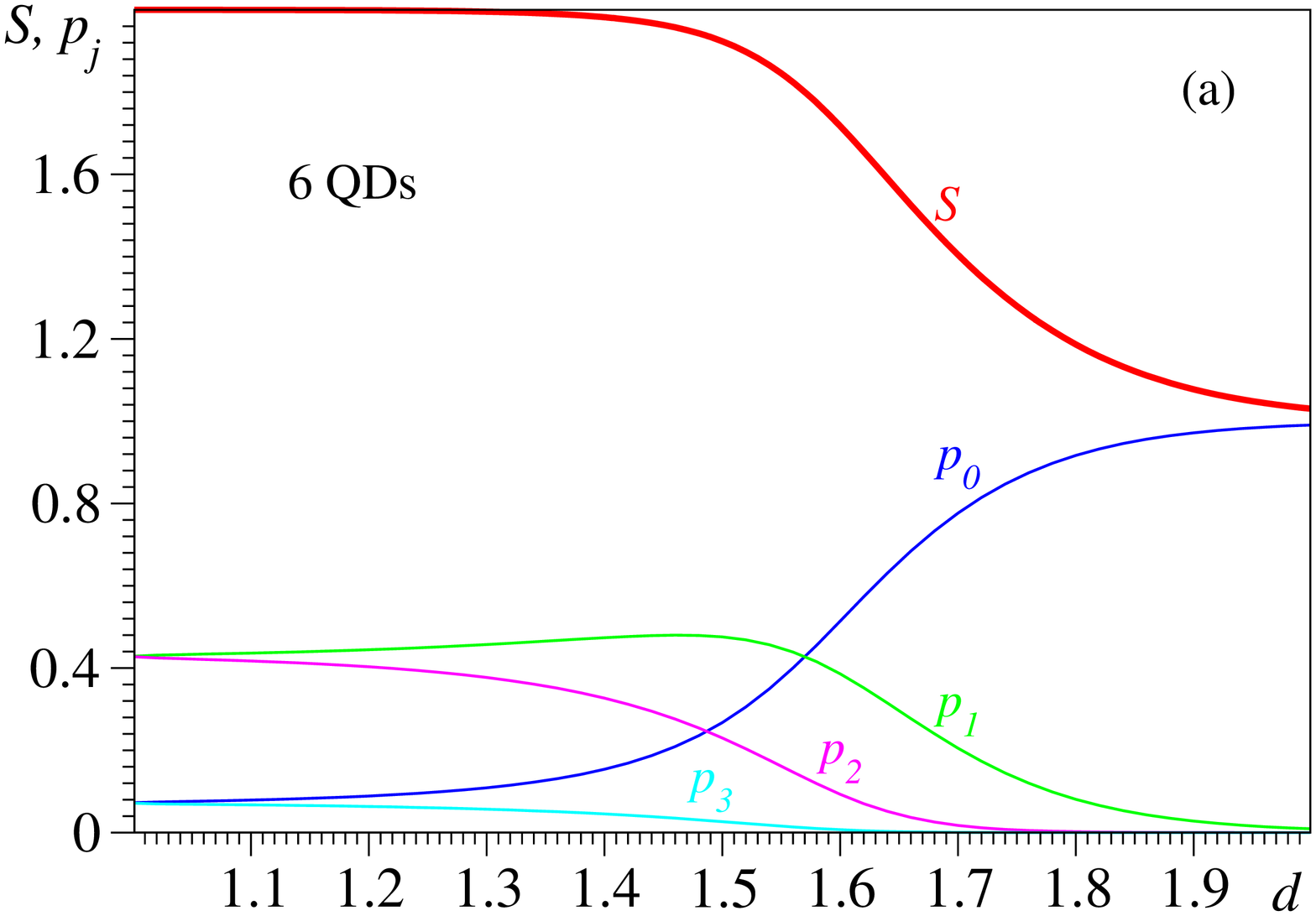}
\includegraphics[width=0.35\textwidth,angle=-90]{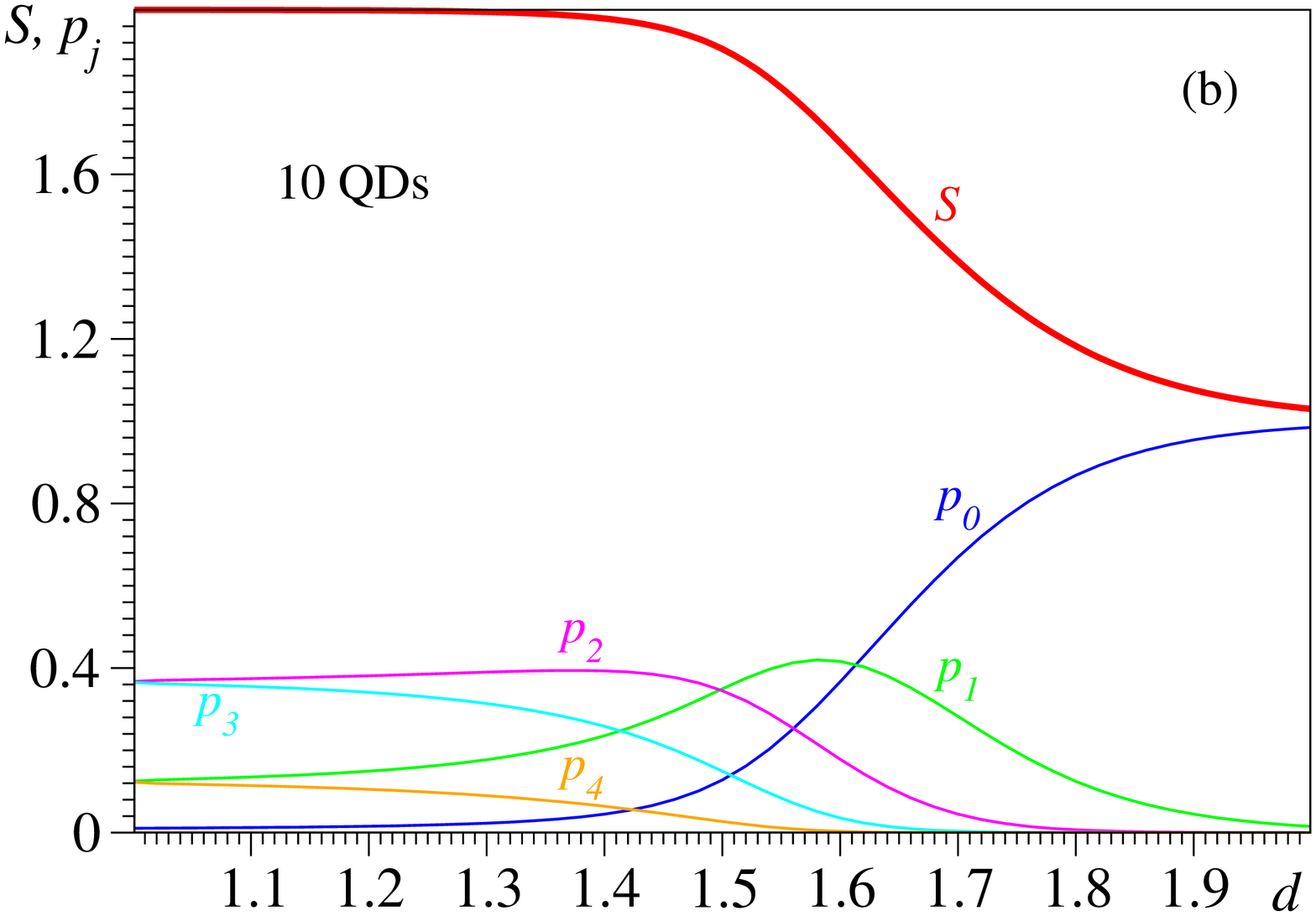}
}
\caption{(Color online) Dependence on the interdot spacing $d$ of the entropy 
($\mathcal{S}$) 
and the total weight $p_0$ , $p_1$, $p_2$  \emph{etc.\ }of the 
multielectronic configurations
in the neutral ground state $\Phi$ of half-filled nanorings with
no, one, two, \emph{etc.\ }doubly occupied dots, respectively.
For the nanoring with ten dots and electrons, the practically vanishing weight of the multielectronic 
configurations with five doubly occupied dots is not shown ($p_5 \simeq 0$).
In both cases, for sufficiently distant QDs, the expansion of the neutral ground state is practically 
exhausted by multielectronic configurations for which none QD is doubly occupied 
($p_0 \approx 1$). The numbers of QDs in the nanoring are specified in the legend.}
\label{fig:S}
\end{figure}

The fact that electron correlations are very important for Ag--QD nanorings was demonstrated 
in a series of previous works. Details can be found in Refs.\ 
\onlinecite{Baldea:2002,Baldea:2004a,Baldea:2007,Baldea:2008} and will be therefore largely omitted here.
One can inspect, \emph{e.\ g.,} the MO occupancies; see Fig.\ 1b of Ref.\ \onlinecite{Baldea:2007} 
and Fig.\ 2 of Ref.\ \onlinecite{Baldea:2008} for ten QD--nanorings. 
At half filling, up to say, $d \alt 1.4$, one can clearly distinguish 
lower, occupied MOs and upper, empty MOs, while towards $d\approx 2$ practically all MOs are 
democratically occupied with half of the number of electrons they can accommodate. 
In the former case, electrons are completely delocalized over the whole nanoring,
whereas in the latter, they are localized on QDs. Rephrasing in a more actual language, 
the quantum entanglement of electron states is nearly perfect 
in the limit of small interdot spacing ($d\agt 1$), 
while electrons become practically disentangled for large interdot spacing ($d \alt 2$). 
The curves for the von Neumann entropy $\mathcal{S}=-\mbox{Tr} \rho \log_{2} \rho $ 
($\rho$ being the reduced density matrix)\cite{BennettEntropy:96,GhirardiEntanglement:04}, 
the quantity used to quantify quantum entanglement,
presented in Fig.\ \ref{fig:S}, nicely visualizes 
how the system smoothly evolves from the limit of perfect entanglement ($\mathcal{S} = 2$) to 
that of perfect disentanglement ($\mathcal{S}=1$) by increasing $d$. For later purposes, 
it is also useful to monitor the weights $p_0$, $p_1$, $p_2$, $\ldots$ 
of the multielectronic configurations in the wave function 
where none or several QDs are doubly occupied ($\sum_i p_i = 1$). 
The weights of these multielectronic configurations 
in the nonionized ground state $\Phi$ are also shown in Fig.\ \ref{fig:S}. Because electron hopping 
dominates at small $d$, one or even several QDs can be doubly occupied. On the contrary, 
on--site Coulomb repulsion $U$ precludes double occupancy at larger $d$, where it is the 
dominant energy term, and all QDs tend to become singly occupied at half filling.
\par
Within the MO--approximation, 
Eq.\ (\ref{eq-w}) immediately yields 
$w_{MO}=1$ for all occupied electron Bloch states. Because Bloch states 
with $+k$ and $-k$ are degenerate for $k\neq 0, N/2$, out of the $N/2$ occupied Bloch states 
(for even $N$) there should be one contribution from the state with $k=0$ and $(N-2)/4$ 
contributions from the states with $\vert k\vert \neq 0$,
i.e., $(N+2)/4$ distinct ionization signals within the MO.
Therefore, to assess the validity of the single--particle picture, we shall 
(i) compare the number of spectral lines in the ionization spectrum with $(N+2)/4$ 
and (ii) monitor the deviations from unity of the quantities $w_{}$.
\section{Results}
\label{sec:results}
In this section, we shall present in detail numerical results for ionization in 
half--filled nanorings consisting of six and ten QDs.
Excepting for Fig.\ \ref{fig:E2-6-QDs}c, in all other cases of Secs.\ 
\ref{sec:6-QDs} and \ref{sec:10-QDs} we only depict the ionization signals 
with significant spectroscopic factors. In addition to these, there are numerous 
spectral lines of very small but still non--vanishing intensities, a fact to be discussed in Sec.\ 
\ref{sec:hidden-symmetry}.
\par
Because of spatial symmetry, free (Bloch) electron states  
satisfy exactly the Hartree--Fock (SCF) equations (see \emph{e.g.,} Ref.\ \onlinecite{fetter}). 
As a consequence, all ionization energies scale as the hopping integral $t_0$ within 
the MO--approximation. 
Therefore, the curves for MO--ionization energies fall off exponentially with $d$. 
In the figures of this section, we shall not draw these rather trivial lines of the MO-approximation. 
We shall present instead the less trivial curves (depicted by black dashed lines
in the panels showing the ionization energies) for the lowest ionization energy $\varepsilon_{loc}$
in the large $d$ limit, where electron hopping is negligible 
($t_0 \to 0$) and none QD is doubly occupied.
In this limit, the energy of the neutral ground state is $N \varepsilon_H + N V$.
Removal of an electron from this state yields an ionized state with the 
energy $(N - 1) \varepsilon_H + (N - 2) V$. The difference of these energies 
represents the lowest ionization energy
$\varepsilon_{loc} = -\varepsilon_H - 2 V$ in the limit of perfect localization.
\par
\begin{figure}[h]
\centerline{
\includegraphics[width=0.35\textwidth,angle=-90]{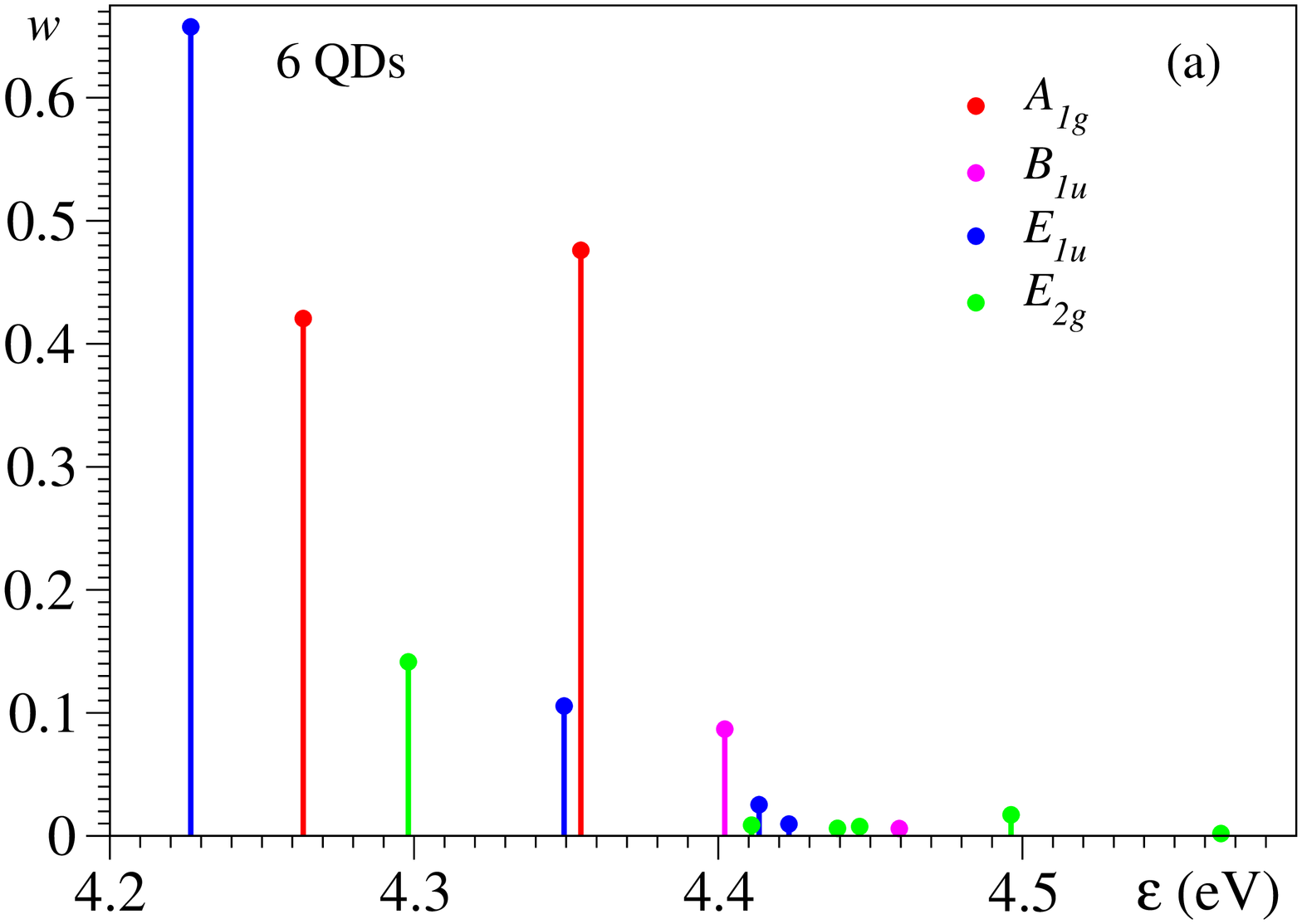}
\includegraphics[width=0.35\textwidth,angle=-90]{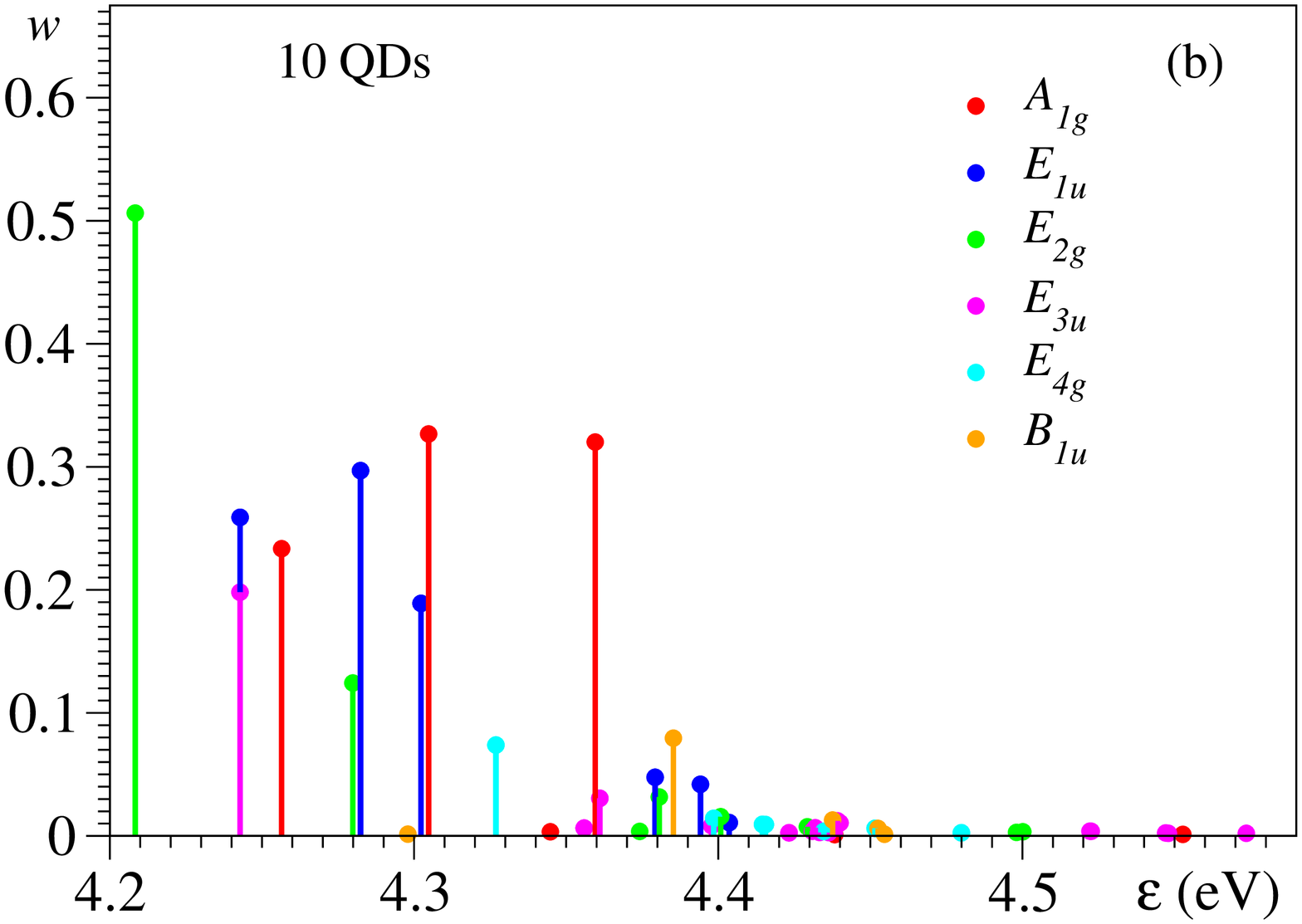}}
\caption{(Color online) Exact ionization spectra for six-- and ten--QD nanorings with 
interdot spacing $d=1.61$.
Notice that, in the MO spectra there exist only two and three lines (all with $w=1$), 
respectively. The MO--lines are located at 4.556\,eV and 4.609\,eV in the former case, 
and at  4.536\,eV,  4.589\,eV, and  4.556\,eV in the latter case.}
\label{fig:full-spectra}
\end{figure}

Before going to discuss correlation effects on ionization spectra in detail, 
to give a flavor on the substantial difference between the MO and exact 
ionization spectra, we present an example in Fig.\ \ref{fig:full-spectra}.
Full spectra for all symmetries are shown in Fig.\ \ref{fig:full-spectra} 
for both six-- and ten--dot nanorings, the two cases we shall later examine at length. 
There and in all of the other cases presented below, we have set 
$\varepsilon_H = -4.504$\,eV, a choice that only fixes the origin for energy 
[see Eq.\ (\ref{eq-model-hamiltonian})].
For a first glance assessment 
of the important role played by correlations for the spectra of Fig.\ \ref{fig:full-spectra} 
it suffices to say that, if the MO picture were valid, the spectra of 
Fig.\ \ref{fig:full-spectra}a and \ref{fig:full-spectra}b 
would only consist of two and three lines, respectively.

\subsection{Six-dot nanorings}
\label{sec:6-QDs}
Before discussing the numerical results for six-dot nanorings 
(point group D$_{6h}$),
it is useful to note that, within the single--particle picture, there are six 
MOs. Ordered by increasing energy these are:
a nondegenerate $a_{1g}$ MO ($k=0$), two degenerate 
$e_{1u}$ ($k=\pm 1$) and $e_{2g}$ ($k=\pm 2$), and a nondegenerate $b_{1u}$ MO ($k=3$).
At half filling the $a_{1g}$-- and $e_{1u}$--MOs are 
occupied by two and four electrons, respectively, whereas $b_{1u}$-- and $e_{2g}$--MOs are empty.
\par
From the multiplication table of the group D$_{6h}$ and because the neutral 
ground state possesses $A_{1g}$ symmetry, the ionization from an MO of a certain symmetry 
will bring the nanoring into a state of the same symmetry, \emph{e.\ g.}, removing an 
electron from an $e_{1u}$--MO yields an ionized $E_{1u}$-state. Below, we shall discuss 
separately the ionization processes, depending on whether the MO 
wherefrom an electron is ejected would be occupied or not within the MO picture. 
\par
\begin{figure}[h]
\centerline{
\includegraphics[width=0.35\textwidth,angle=-90]{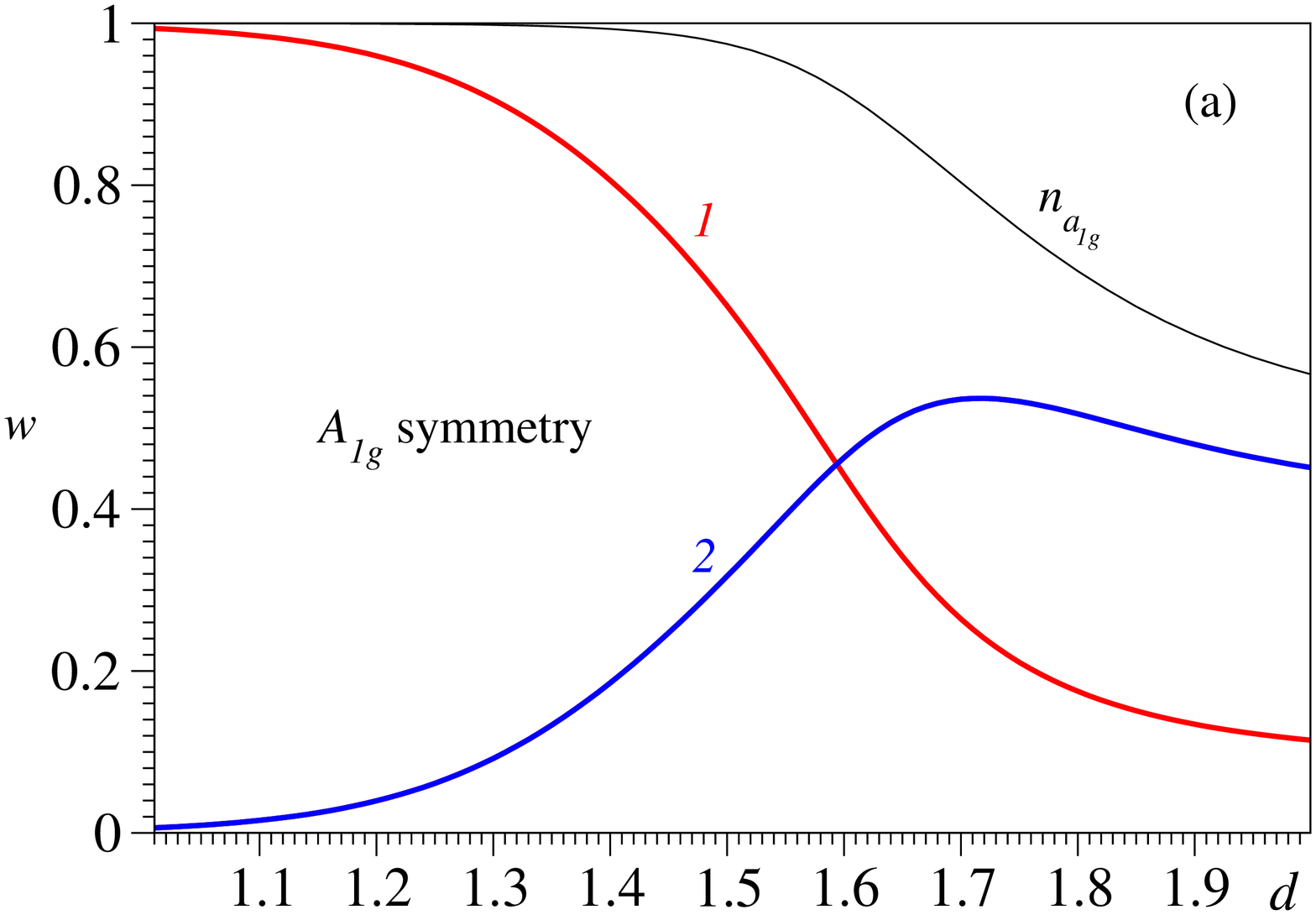}
\includegraphics[width=0.35\textwidth,angle=-90]{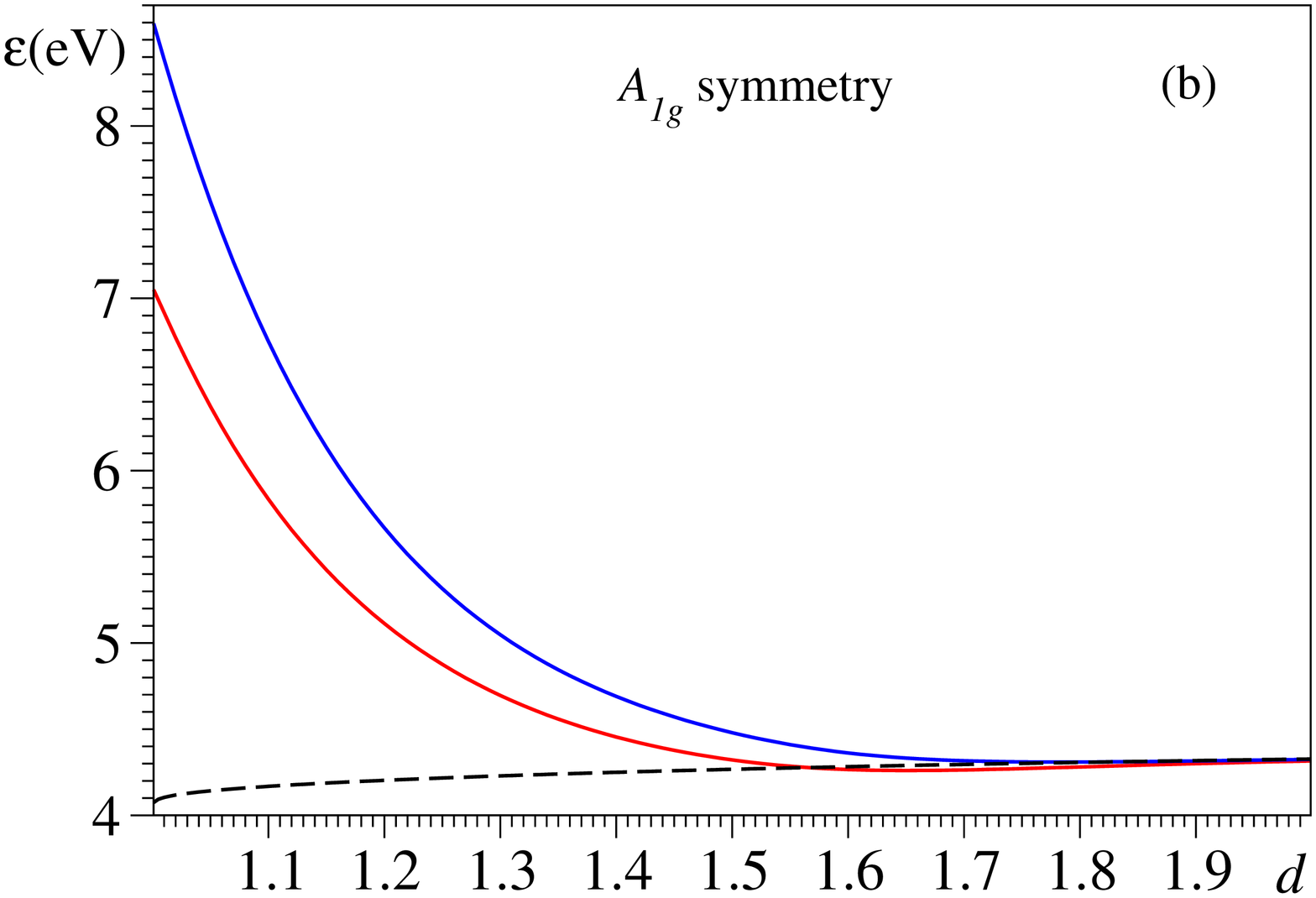}}
\centerline{
\includegraphics[width=0.35\textwidth,angle=-90]{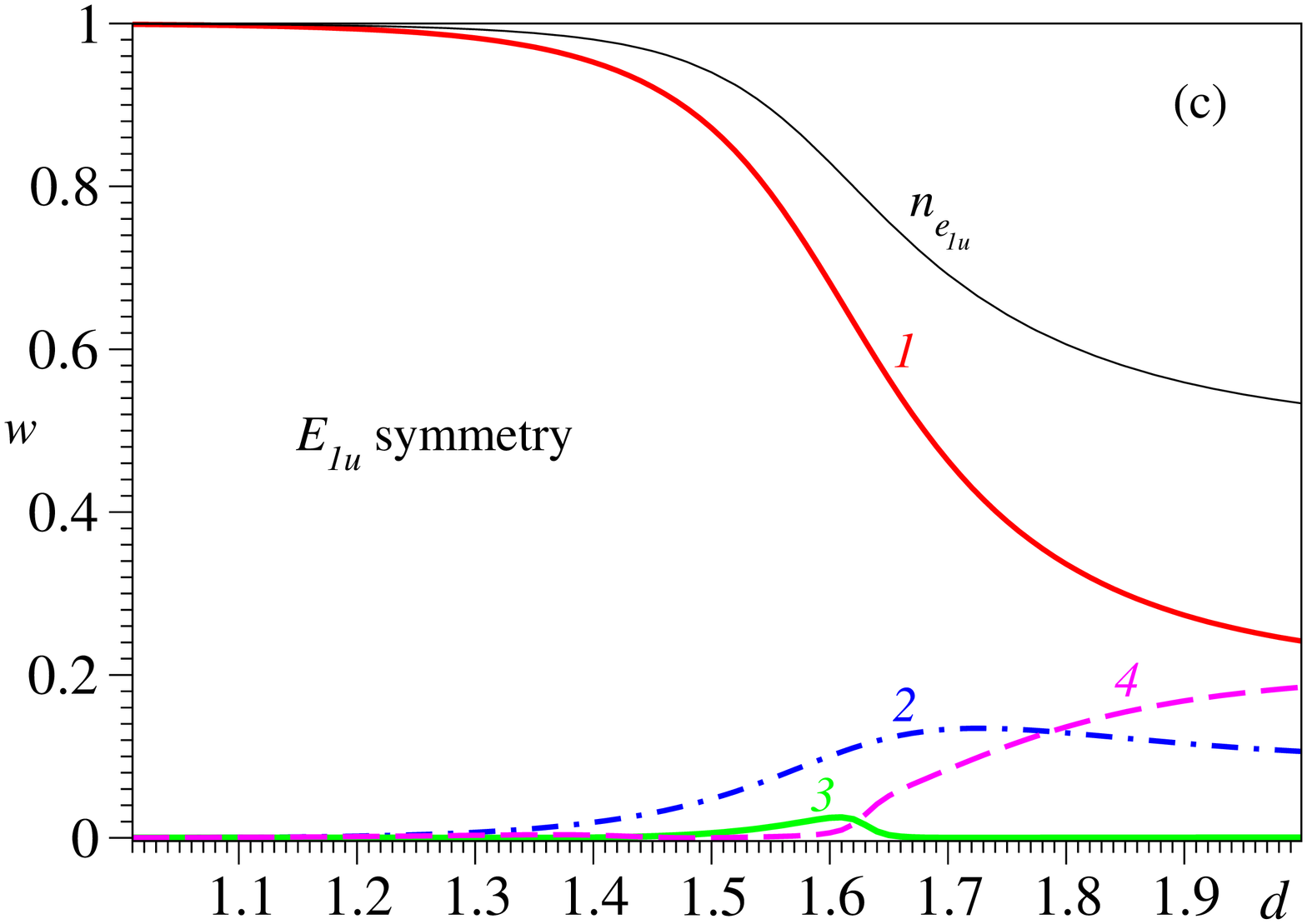}
\includegraphics[width=0.35\textwidth,angle=-90]{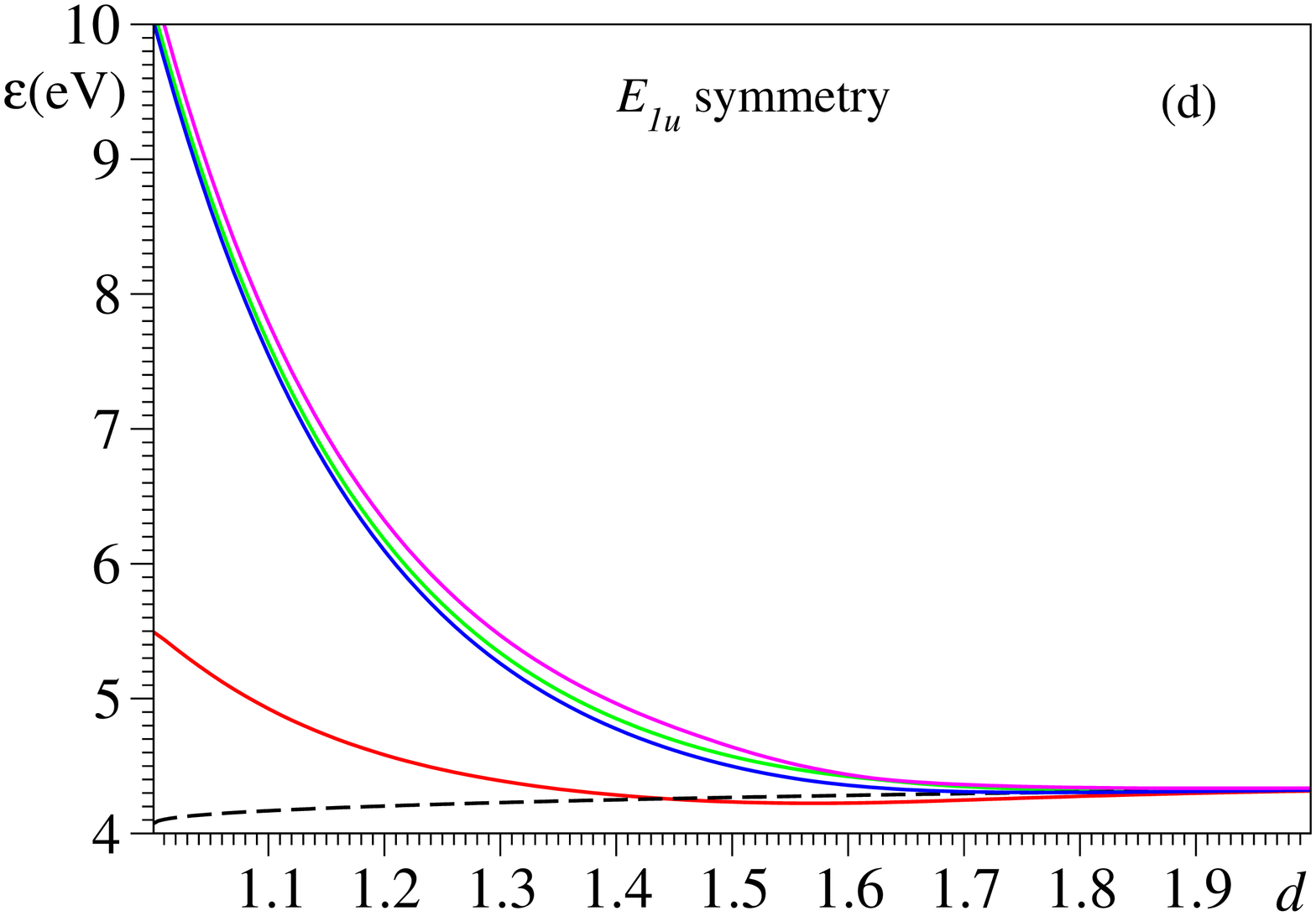}}
\caption{(Color online) $A_{1g}$-- and $E_{1u}$--spectral factors $w$ 
and ionization energies $\varepsilon$ versus interdot spacing $d$ in    
six--QD nanorings. The numbers $i=1, 2,\ldots$ in the legend 
label the ionized eigenstates $\Psi_{k,i}$ [\emph{cf.\ }Eq.\ (\ref{eq-w})].
In panels (b) and (d), the black dashed line corresponds to the lowest ionization 
energy $ - \varepsilon_{H} - 2 V$ in the limit of perfect localization.}
\label{fig:A1-E1-6-QDs}
\end{figure}

Within the MO picture only two ionization processes are possible, 
because only the $a_{1g}$-- and $e_{1u}$--MOs are occupied. Accordingly, only 
two ionization processes (denoted by $1$ in 
Figs.\ \ref{fig:A1-E1-6-QDs}a and \ref{fig:A1-E1-6-QDs}c, respectively) 
are significant at small $d$.
The latter, which corresponds to the lowest ionization energy (HOMO-ionization), 
brings the nanoring into the ionized ground state. 
It is denoted by $1$ in Fig.\ \ref{fig:A1-E1-6-QDs}c and was already 
considered in Ref.\ \onlinecite{Baldea:2002}.
With increasing $d$, the spectroscopic factors of these processes decrease from the MO--value $w=1$.
This represents one manifestation of electron correlations. 
Another manifestation is the fact that they become even significantly smaller than 
the corresponding MO occupancies: $w_{k,1} < n_{k}$ ($k=0, 1$). 
(Henceforth we omitt the spin label whenever unncessary and write, 
\emph{e.~g.}, $n_{k}$ instead of $n_{k,\sigma}$.)
Compare the curves denoted by $1$ with the 
MO populations $n_{0}\equiv n_{A_{1g}}$ and $n_{1}\equiv n_{E_{1u}}$
in Figs.\ \ref{fig:A1-E1-6-QDs}a and \ref{fig:A1-E1-6-QDs}c,
respectively. According to the sum rule (\ref{eq-sum-rule}), other 
eigenstates should aquire significant spectroscopic factors, and this is 
indeed what one observes in Figs.\ \ref{fig:A1-E1-6-QDs}a and \ref{fig:A1-E1-6-QDs}c.
The differences $1 - n_{1, 0}$ can be considered a measure of the extent to which 
electron correlations affect the $e_{1u}$-- and $a_{1g}$--MOs of the neutral ground state.
In comparison with the $A_{1g}$--ionization spectrum, 
where two ionized $A_{1g}$--eigenstates practically exhaust all 
spectral weight in the whole $d$-range of interest, that for $E_{1u}$--symmetry is somewhat 
richer; compare Fig.\ \ref{fig:A1-E1-6-QDs}a with Fig.\ \ref{fig:A1-E1-6-QDs}c. 
In the latter case we encounter 
an avoided crossing at $d = 1.633$, where the energy difference between the curves denoted by 
$4$ and $3$ attains its minimum values of $28.7$\,meV. 
Later on in this paper we shall return to a more detailed analysis 
of the phenomenon of avoided crossing.
Here we only note that, by considering the diabatic approximation for the states participating to 
this avoided crossing, basically only three (diabatic) states give nonvanishing contributions 
to the $E_{1u}$--spectrum. 
\par
\begin{figure}[h]
\centerline{
\includegraphics[width=0.35\textwidth,angle=-90]{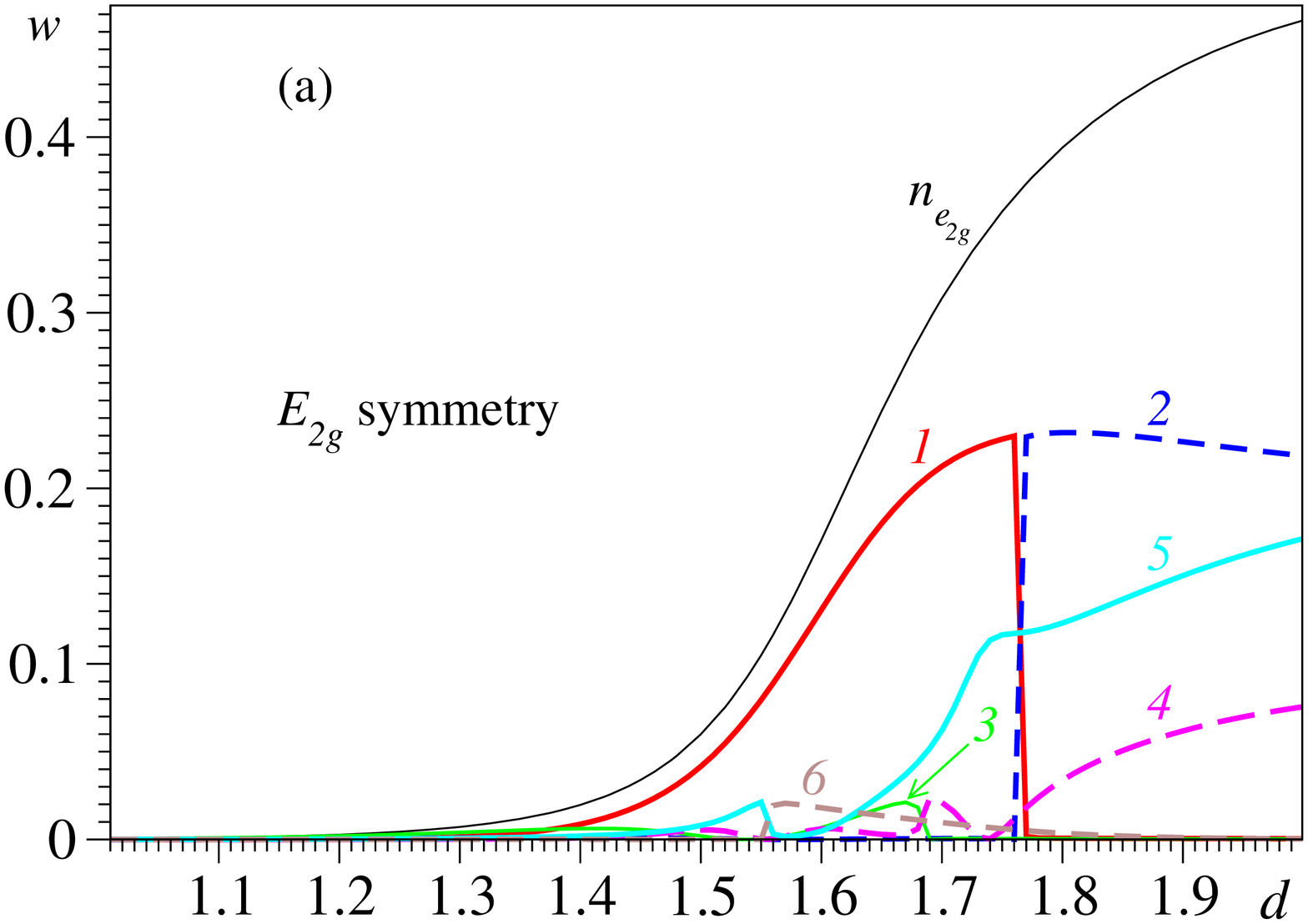}
\includegraphics[width=0.35\textwidth,angle=-90]{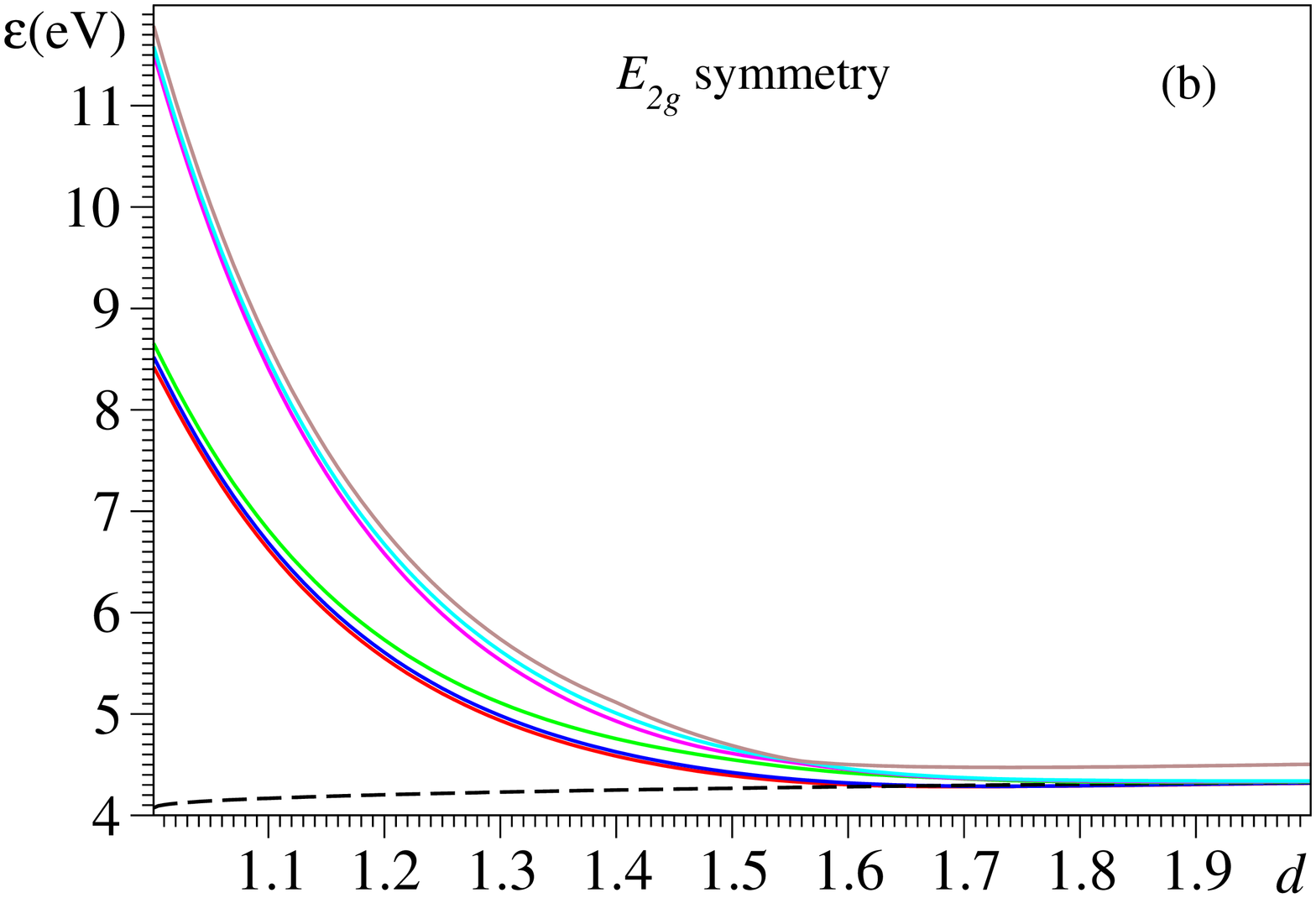}
}
\centerline{\hspace*{-0ex}
\includegraphics[width=0.35\textwidth,angle=-90]{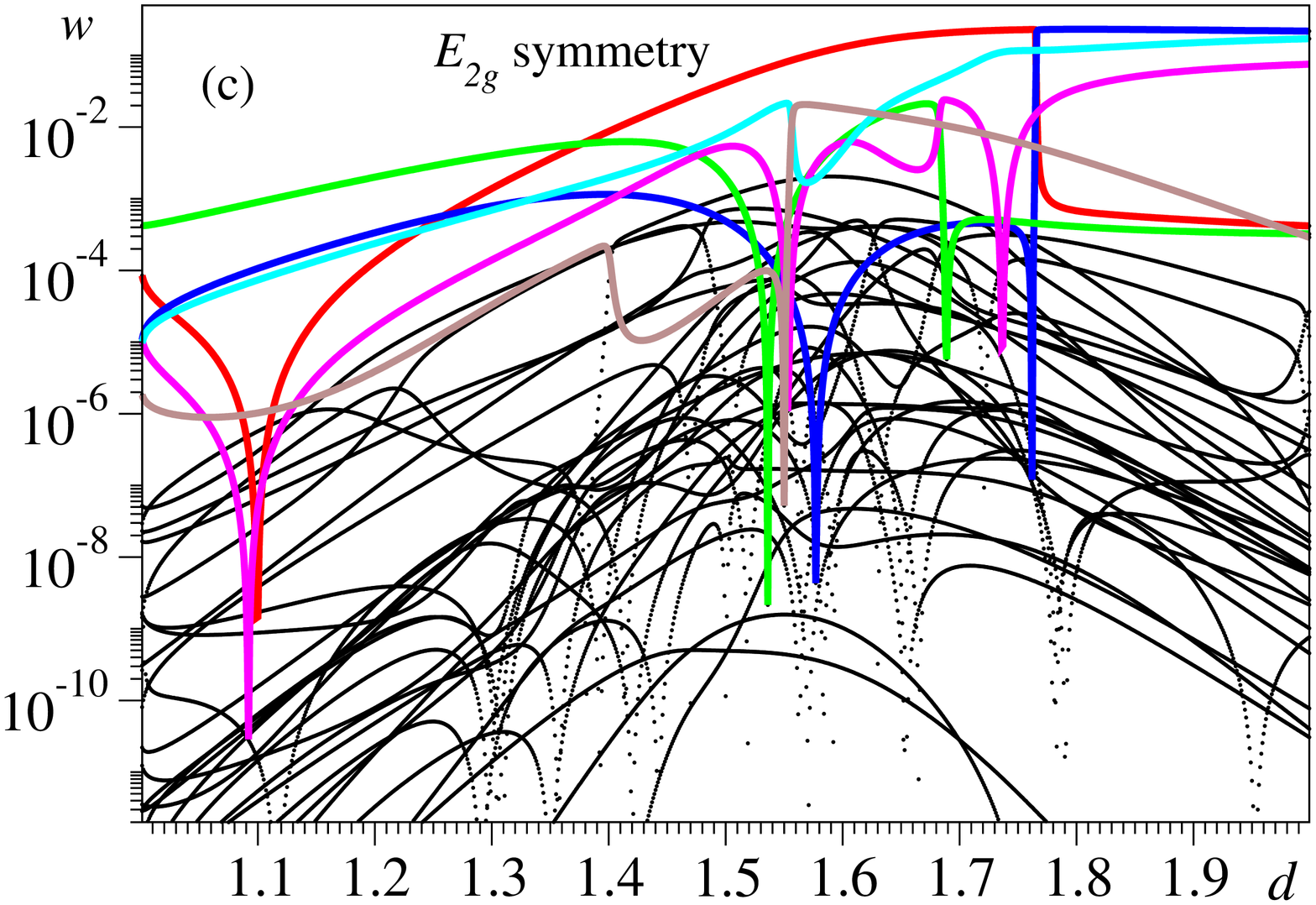}
}
\caption{(Color online) $E_{2g}$--spectral factors $w$ (a) and ionization energies $\varepsilon$ (b)
versus interdot spacing $d$ in    
six--QD nanorings. The numbers $i=1,2,\ldots$ in the legend 
label the ionized eigenstates $\Psi_{k,i}$ [\emph{cf.\ }Eq.\ (\ref{eq-w})].
In panel (b), the black dashed line corresponds to the lowest ionization 
energy $ - \varepsilon_{H} - 2 V$ in the limit of perfect localization.
Notice the numerous spectral lines with very small intensities in panel (c), 
which represents panel (a) redrawn using 
the logarithmic scale on the ordinate, demonstrating that the extended Hubbard model 
is characterized by a hidden \emph{quasi}--symmetry (see Sec.\ \ref{sec:hidden-symmetry}.)}
\label{fig:E2-6-QDs}
\end{figure}
\begin{figure}[h]
\centerline{
\includegraphics[width=0.35\textwidth,angle=-90]{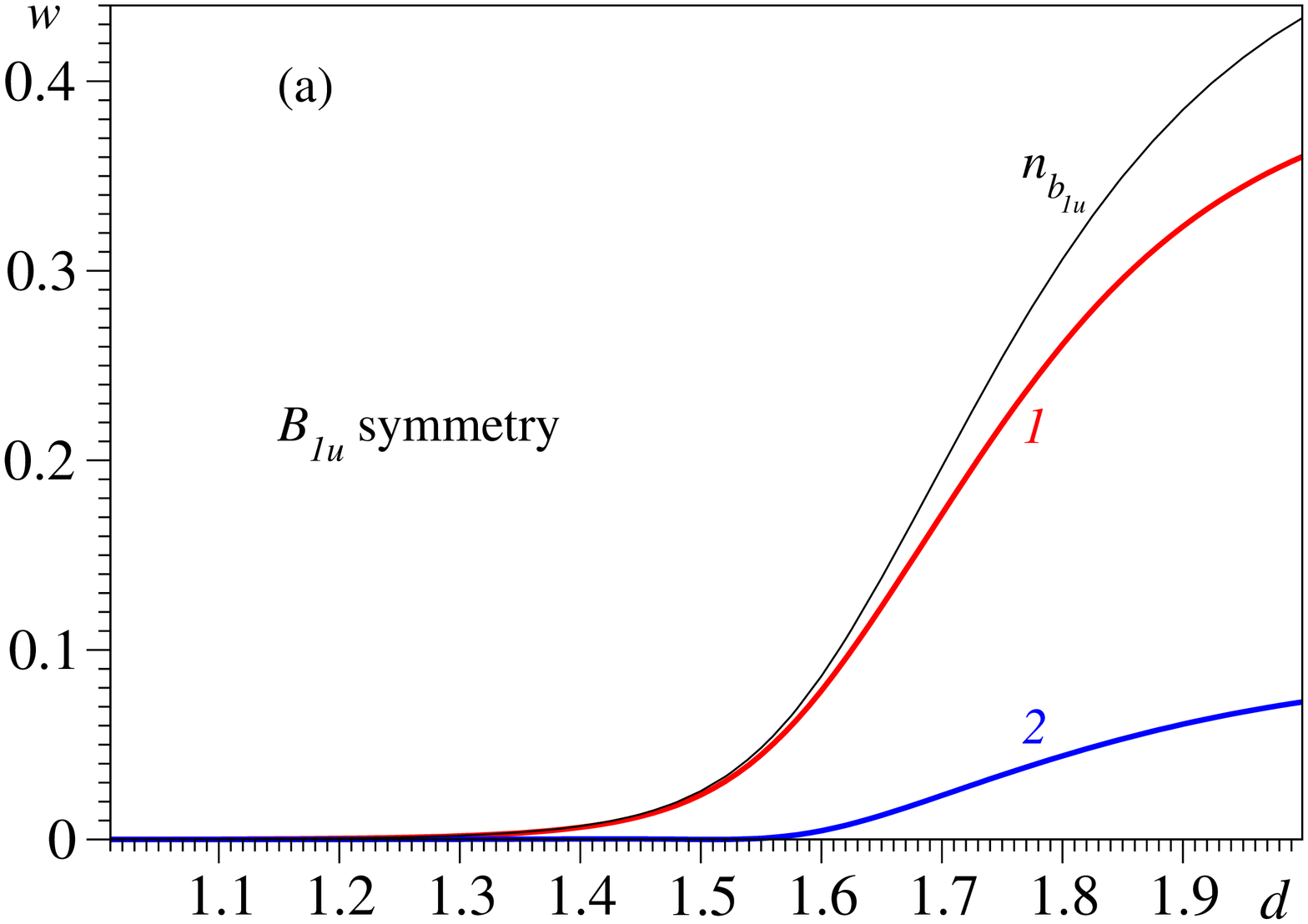}
\includegraphics[width=0.35\textwidth,angle=-90]{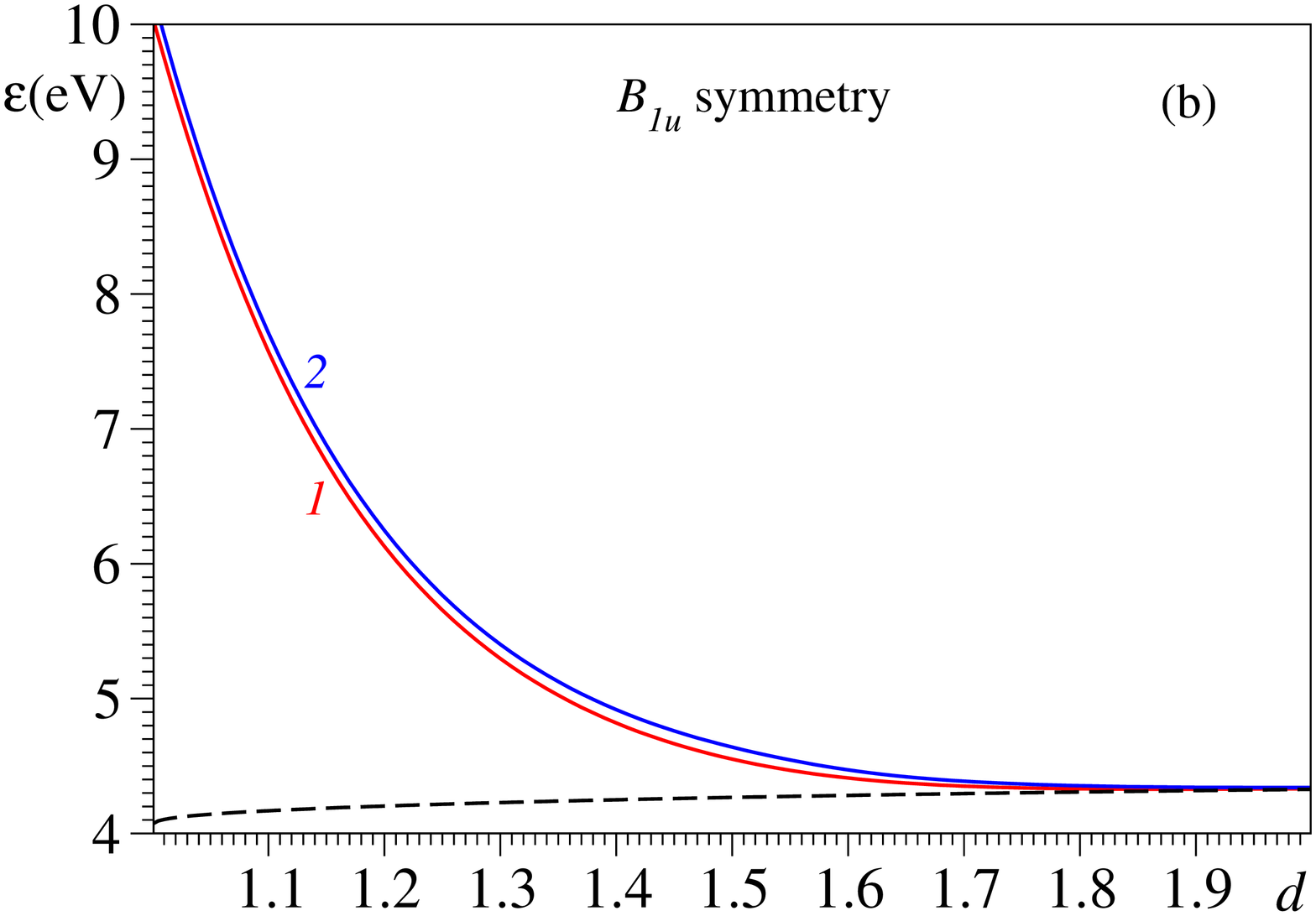}
}
\caption{(Color online) $B_{1u}$--spectral factors $w$ (a) and ionization energies $\varepsilon$ (b)
versus interdot spacing $d$ in    
six--QD nanorings. The numbers $i=1$ and $2$ in the legend 
label the ionized eigenstates $\Psi_{k,i}$ [\emph{cf.\ }Eq.\ (\ref{eq-w})].
In panel (b), the black dashed line corresponds to the lowest ionization 
energy $ - \varepsilon_{H} - 2 V$ in the limit of perfect localization.}
\label{fig:B1-6-QDs}
\end{figure}

Let us now switch to ionization from MOs that would be empty 
(in our case, $e_{2g}$ and $b_{1u}$) if the MO picture were valid.
Such ionization processes are due to ground state 
correlations \cite{Cederbaum:1986}; in the absence of the latter 
they do not obviously exist. Our results for these ionization processes are collected in 
Figs.\ \ref{fig:E2-6-QDs} and \ref{fig:B1-6-QDs}. In accord with the fact that 
electron correlations in the neutral ground state increase with $d$, 
a general trend of increasing spectroscopic factors is visible in these figures. 
Particularly interesting is the $E_{2g}$--ionization spectrum, where a 
series of avoided crossings can be seen. 
As visible in Figs.\ \ref{fig:E2-6-QDs}a, one avoided crossing for each of 
the pairs of states (1, 2), (3, 4), and (5, 6) occurs 
at $d\simeq 1.765$, $d\simeq 1.685$, and $d\simeq 1.555$, respectively. 
The corresponding minimum values of the energy differences 
amount $0.16$\,meV, $1.0$\,meV, and $3.1$\,meV, respectively.
In fact, in the avoided crossing at $d\simeq 1.555$, not only two, but rather three 
(namely, 4, 5, and 6) states are involved. This fact is invisible in Fig.\ \ref{fig:E2-6-QDs}a,
because the spectroscopic factor of the state 4 in this region is insignificant.
In comparison with the $E_{2g}$--spectrum, the spectroscopic factors of the 
$B_{1u}$--spectrum are smaller: compare Fig.\ \ref{fig:B1-6-QDs}a with 
\ref{fig:E2-6-QDs}a. This agrees with the fact that 
the lower the unoccupied MO, the more it becomes populated due to the 
ground state correlations.   
\par
Actually, the states targeted by ionization in the case considered in this section 
represent eigenstates of a six--QD nanoring with five electrons. 
In this nanosystem, as discussed recently \cite{Baldea:2008}, optical absorption 
allows to target eigenstates of $A_{1g}$, $A_{2g}$ and $E_{2g}$ symmetries.
So, ionization and optical absorption provide complementary information. 
Some states can be studied only by one method 
($E_{1u}$ and $B_{1u}$ states only by ionization, $A_{2g}$ states 
only by optical absorption), while other states ($A_{1g}$ and $E_{2g}$) by both methods.
For the latter, the information extracted from optical absorption can be compared with 
that from ionization. By comparing the ionization and optical absorption spectra for $A_{1g}$-symmetry
(the present Fig.\ \ref{fig:A1-E1-6-QDs}a and Fig.\ 6f of Ref.\ \onlinecite{Baldea:2008}, 
noting the logarithmic scale in the latter), one can conclude that, in both cases, 
only the first two $A_{1g}$-states give an important contribution. 
Likewise, the same first six $E_{2g}$-states are the only ones that are important 
both for ionization and for optical absorption: compare 
the present Fig.\ \ref{fig:E2-6-QDs}a with Fig.\ 6h of Ref.\ \onlinecite{Baldea:2008}.
Still, all the three avoiding crossings at $d\simeq 1.765$, $d\simeq 1.685$, 
and $d\simeq 1.555$ are visible 
in both spectra. The only difference is the participation in the optical absorption 
of three states at the avoided crossings at 
$d\simeq 1.685$ and $d\simeq 1.555$ (3, 4 and 5 in the former case, and 4, 5, and 6 in the latter),
which is invisible in the $E_{2g}$--ionization spectrum. This is due to the 
insignificant spectroscopic factors of the states 3 and 4 in the corresponding regions
(see Fig.\ \ref{fig:E2-6-QDs}a).

\subsection{Ten-dot nanorings}
\label{sec:10-QDs}
For rings with ten QDs, within the single--particle picture there are ten MOs.
Ordered by increasing energy, they are: a nondegenerate $a_{1g}$ MO ($k=0$), 
two degenerate 
$e_{1u}$ ($k=\pm 1$), $e_{2g}$ ($k=\pm 2$), $e_{3u}$ ($k=\pm 3$), $e_{4g}$ ($k=\pm 4$) 
MOs, and a nondegenerate $b_{1u}$ MO ($k=5$). 
Similar to the case analyzed in Sec.\ \ref{sec:6-QDs}, the ejection of an electron from an 
MO of a given symmetry leads to a final ionized state of the same symmetry, 
\emph{e.\ g.}, removing an electron from an $e_{4g}$--MO yields an ionized $E_{4g}$--state.
This follows from the properties of the D$_{10,h}$ point group and the fact 
that the neutral ground state possesses $A_{1g}$ symmetry.
\par
\begin{figure}[h]
\centerline{
\includegraphics[width=0.35\textwidth,angle=-90]{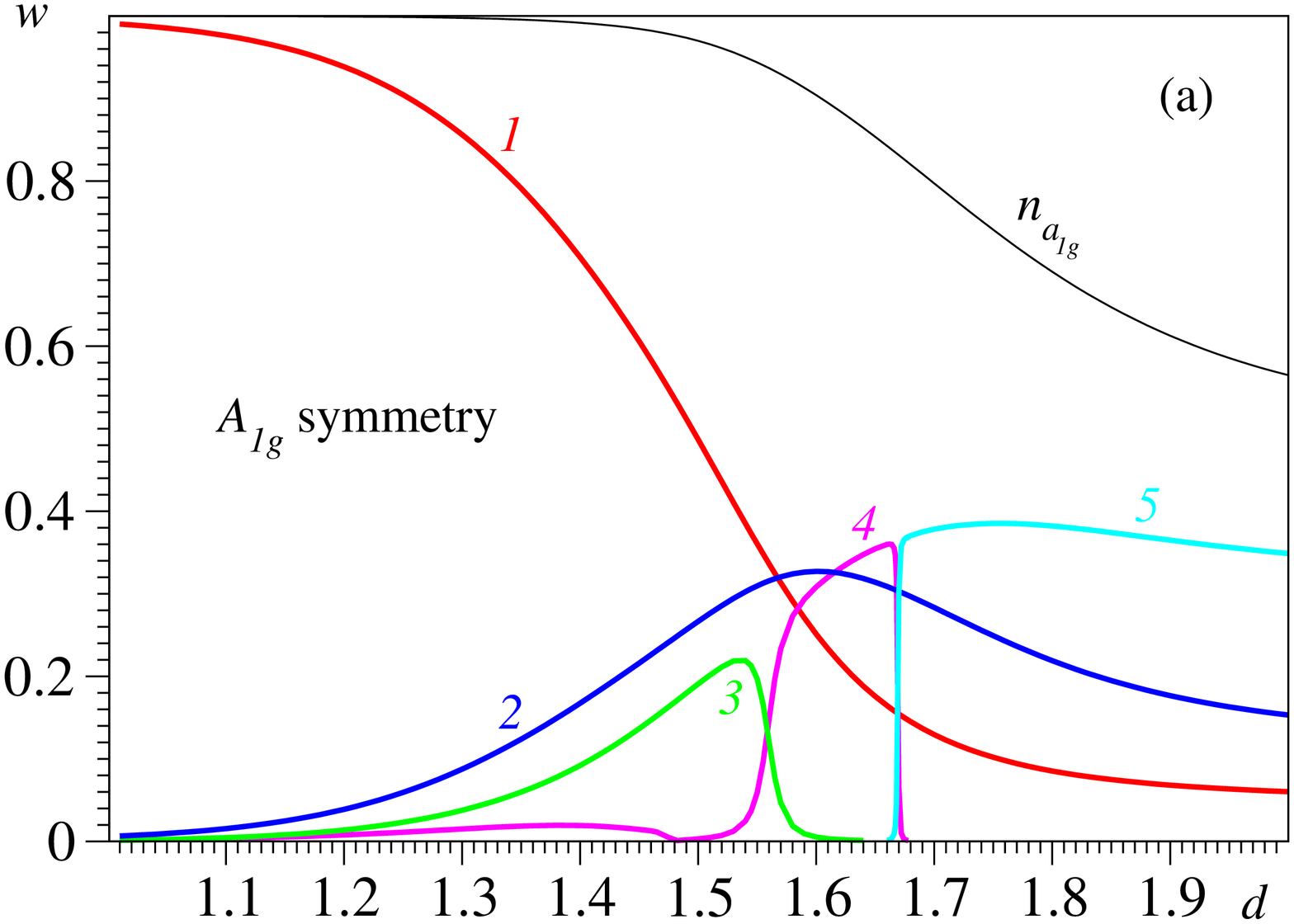}
\includegraphics[width=0.35\textwidth,angle=-90]{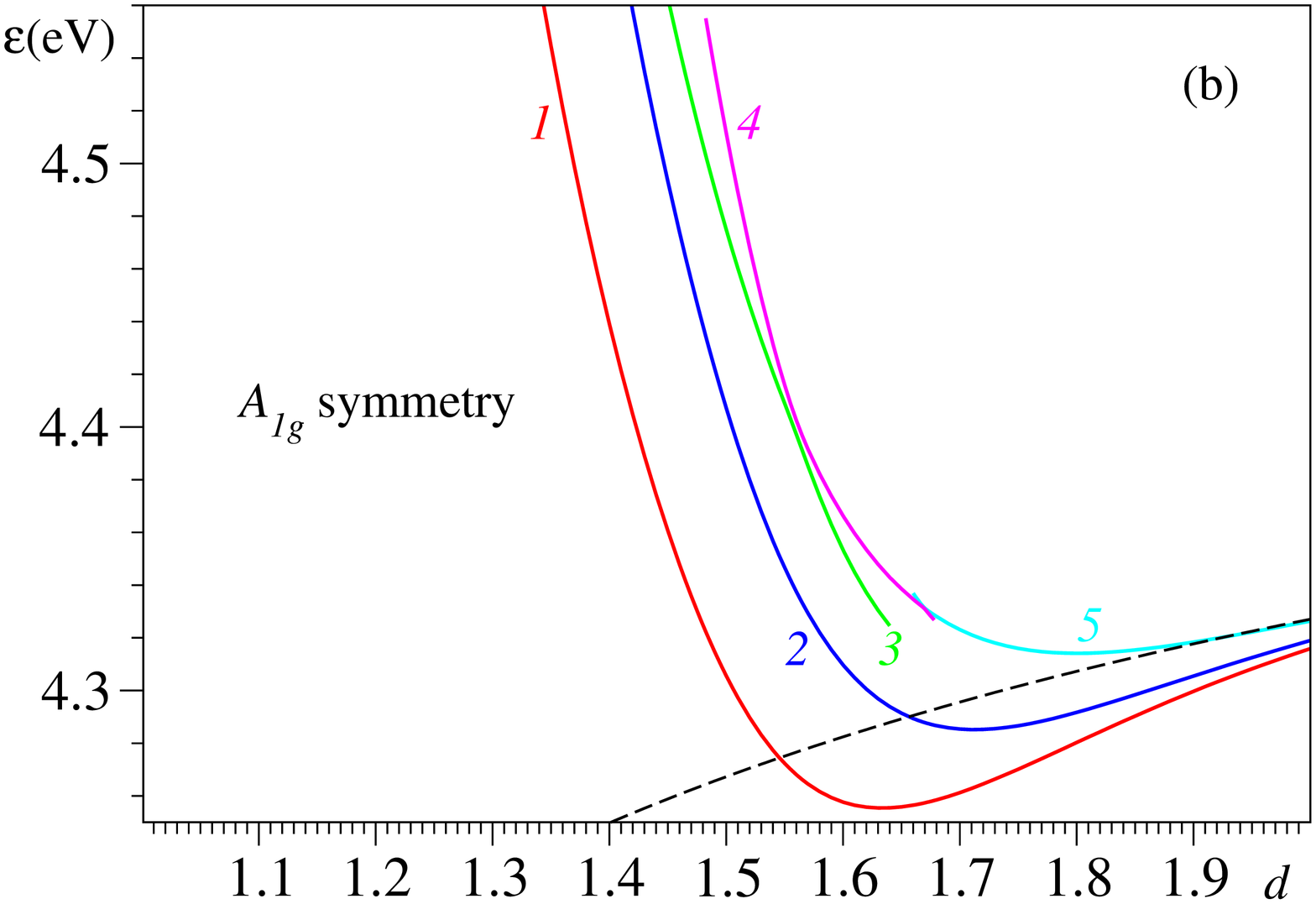}
}
\centerline{
\includegraphics[width=0.35\textwidth,angle=-90]{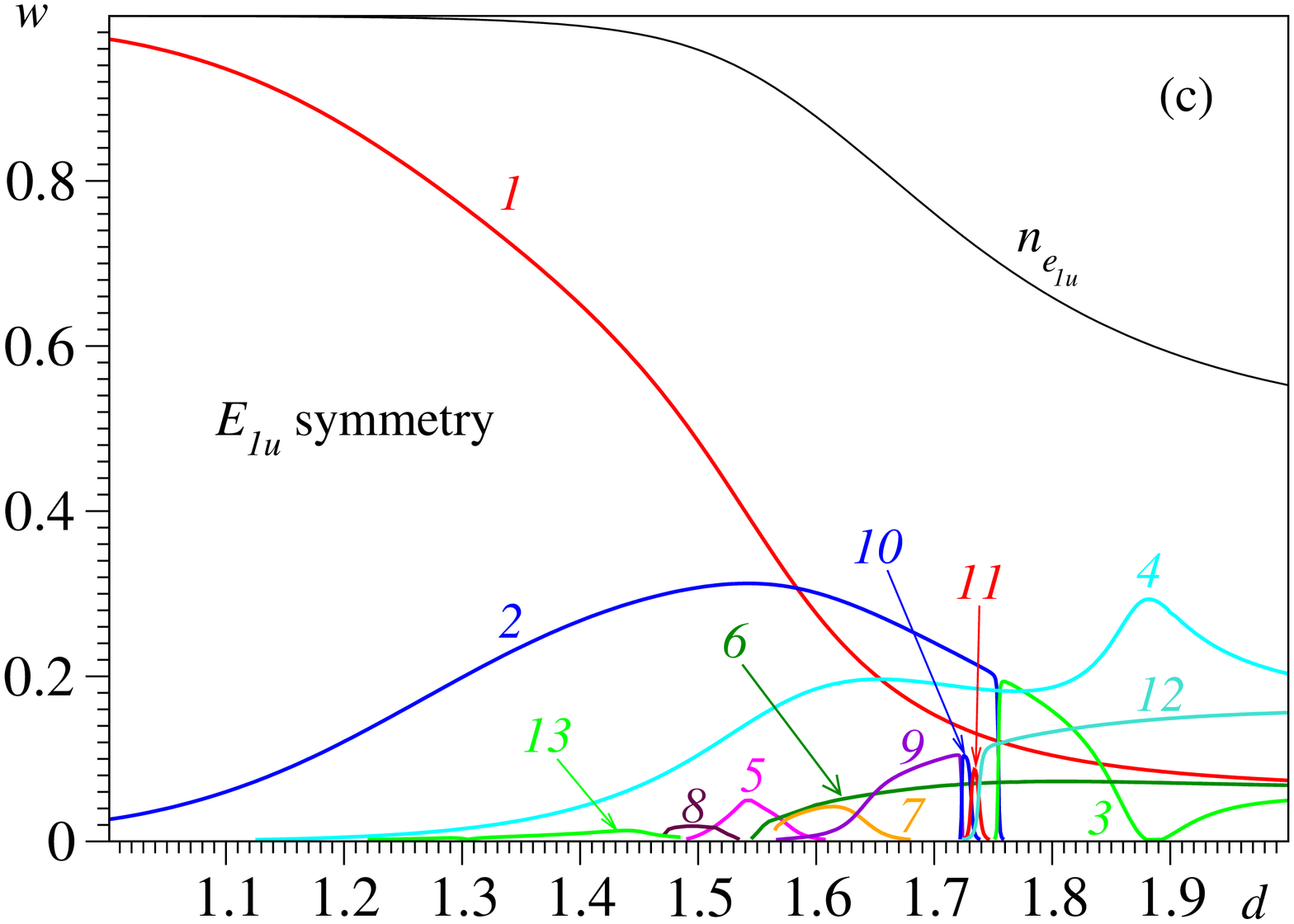}
\includegraphics[width=0.35\textwidth,angle=-90]{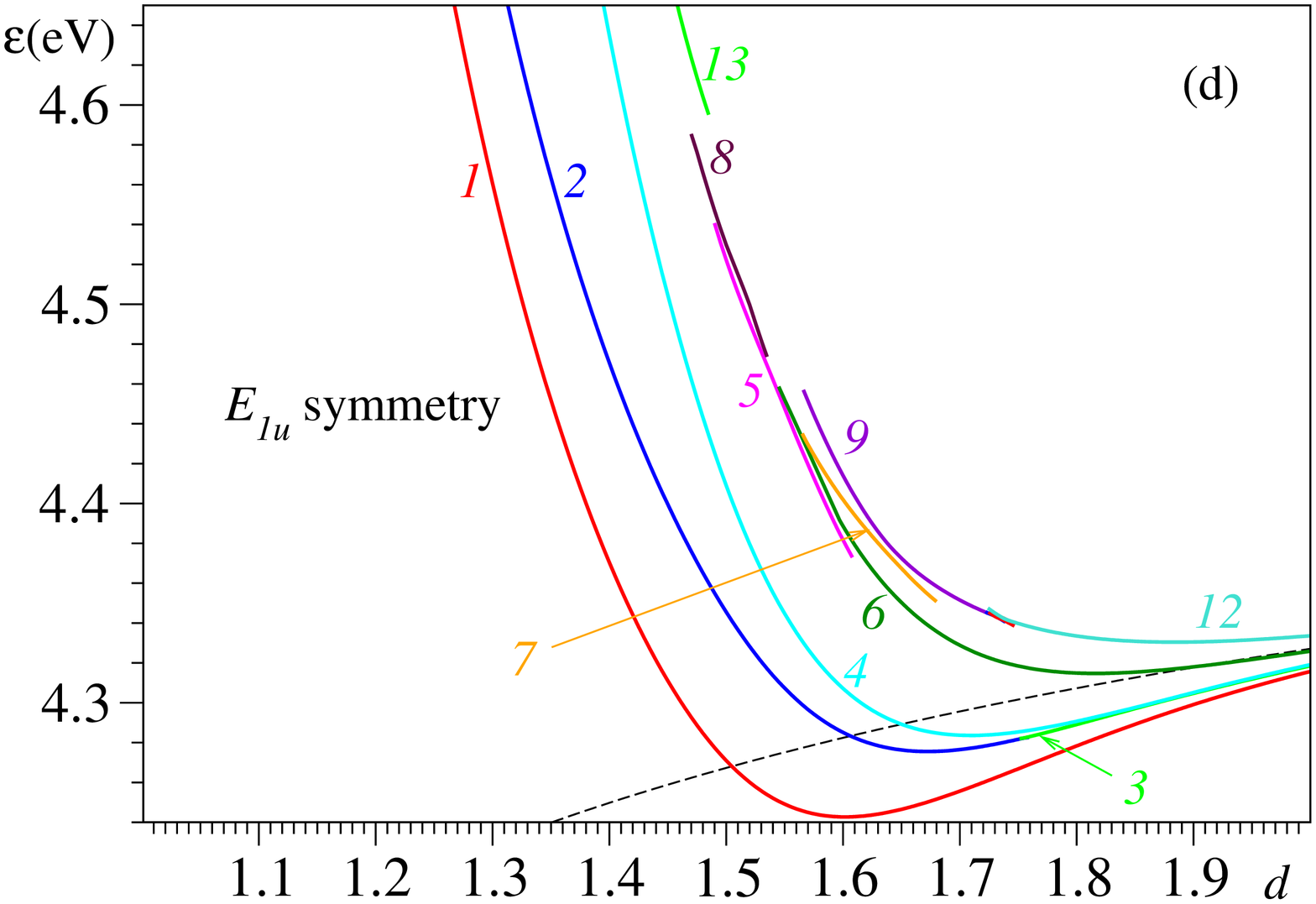}
}
\centerline{
\includegraphics[width=0.35\textwidth,angle=-90]{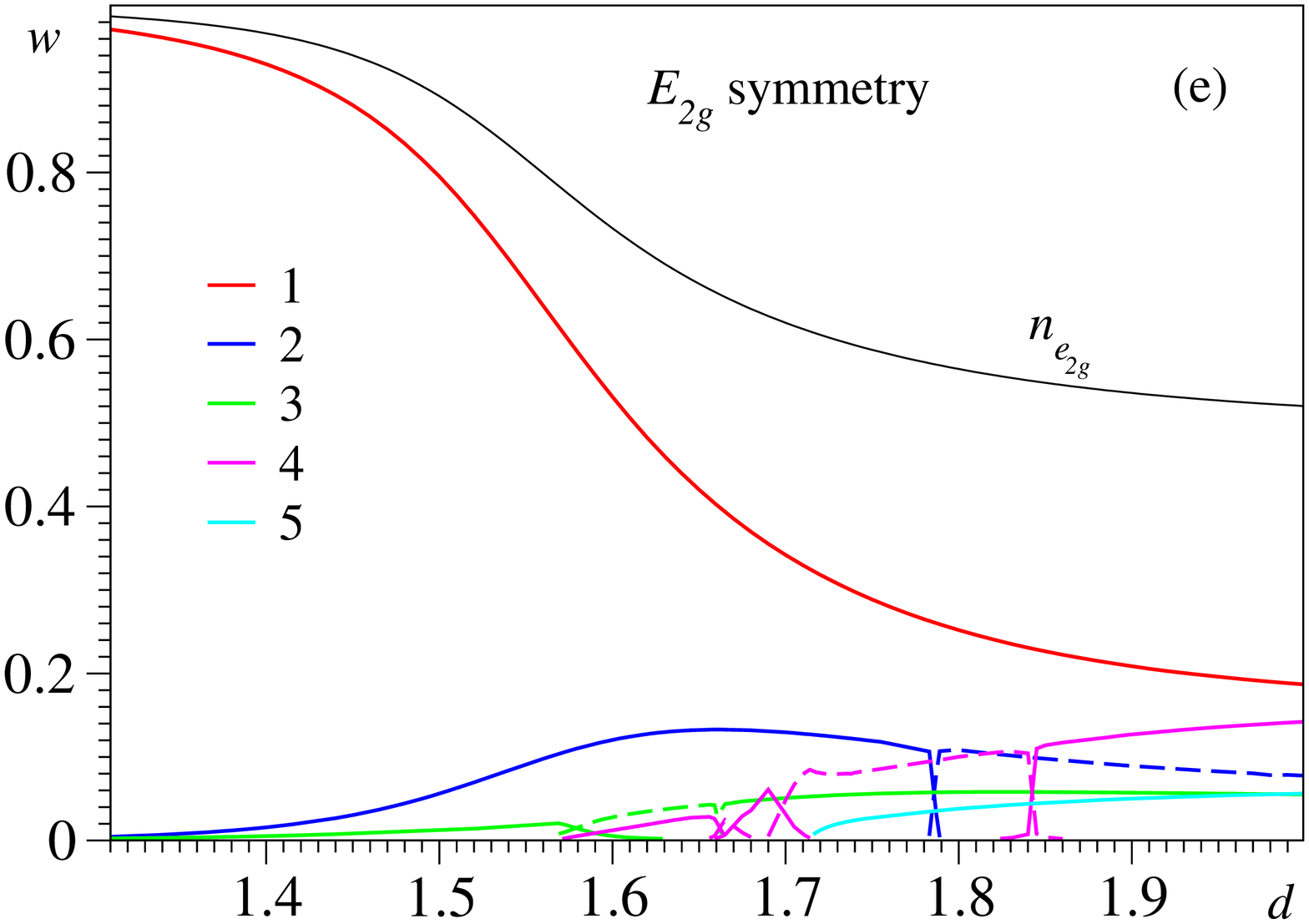}
\includegraphics[width=0.35\textwidth,angle=-90]{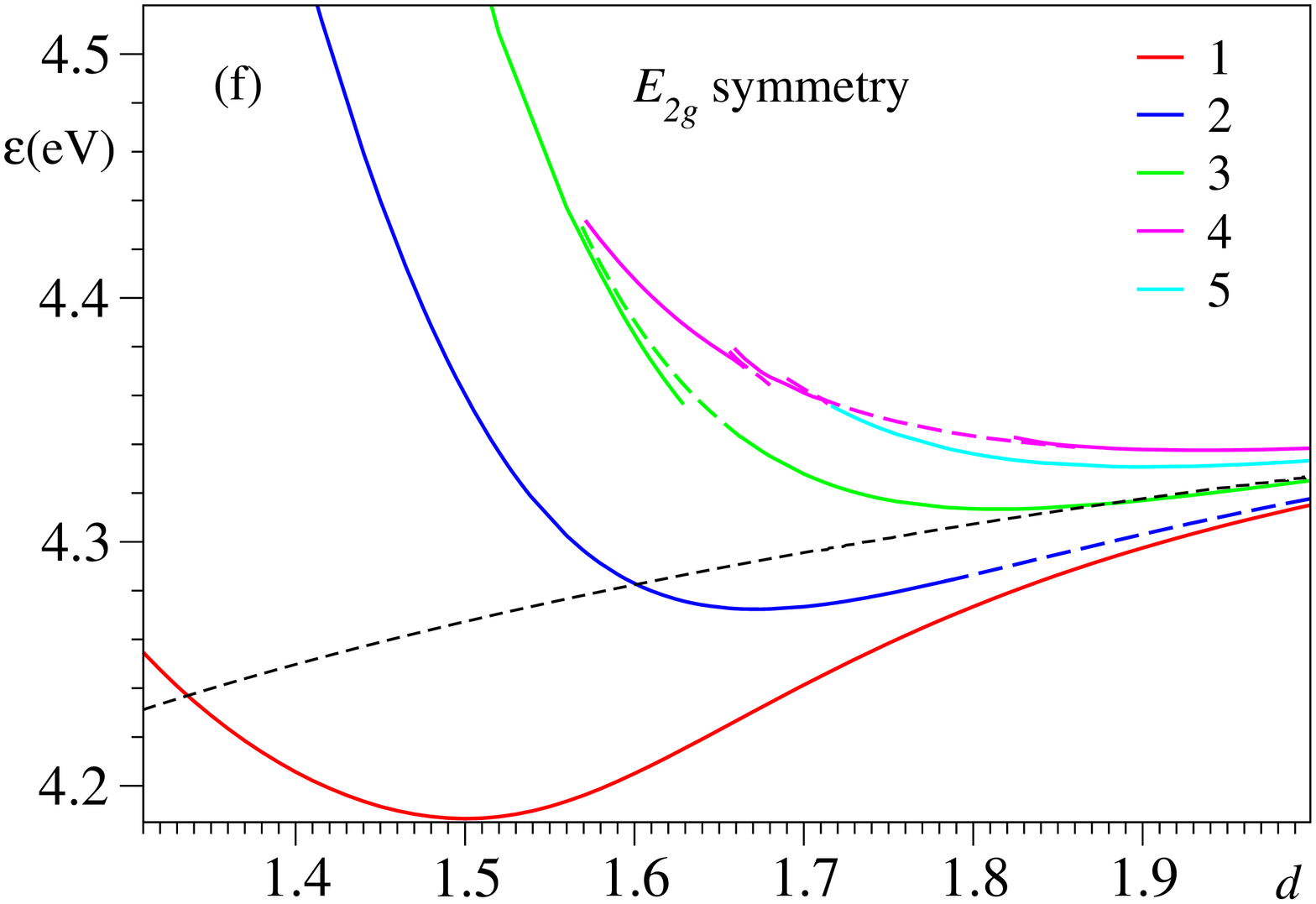}
}
\caption{(Color online) $A_{1g}$--, $E_{1u}$--, and $E_{2g}$--spectral factors $w$ 
and ionization energies $\varepsilon$ 
versus interdot spacing $d$ in ten--QD nanorings. The numbers $i=1,2,\ldots$ in the legend 
label the ionized eigenstates $\Psi_{k,i}$ [\emph{cf.\ }Eq.\ (\ref{eq-w})].
In panel (d), the states 10 and 11 cannot be distinguished from the curves for 9 and 12 
within the drawing accuracy.
In panels (b), (d), and (f), the black dashed line corresponds to the lowest ionization 
energy $ - \varepsilon_{H} - 2 V$ in the limit of perfect localization.}
\label{fig:A1-E1-E2-10-QDs}
\end{figure}
\begin{figure}[h]
\centerline{
\includegraphics[width=0.35\textwidth,angle=-90]{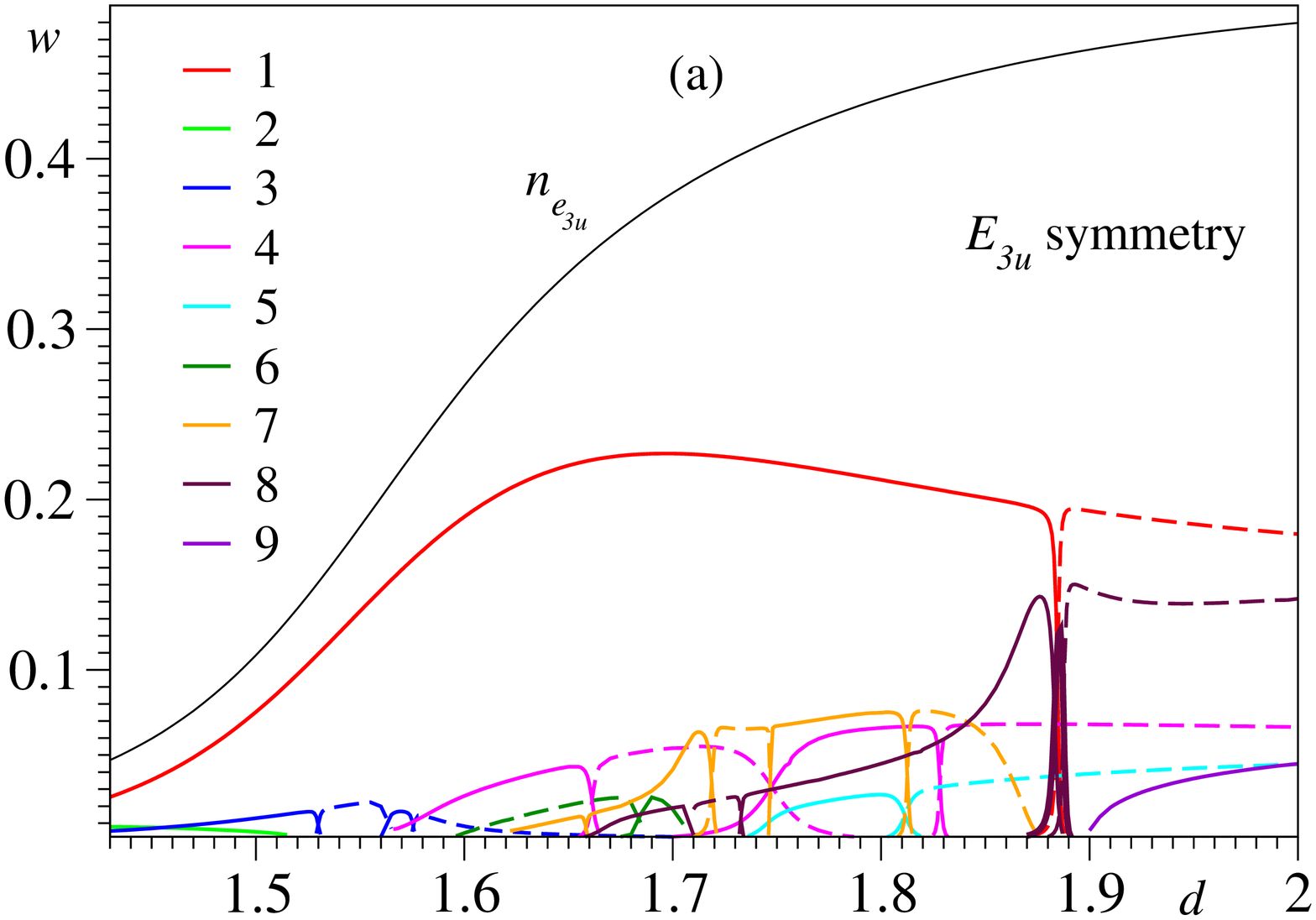}
\includegraphics[width=0.35\textwidth,angle=-90]{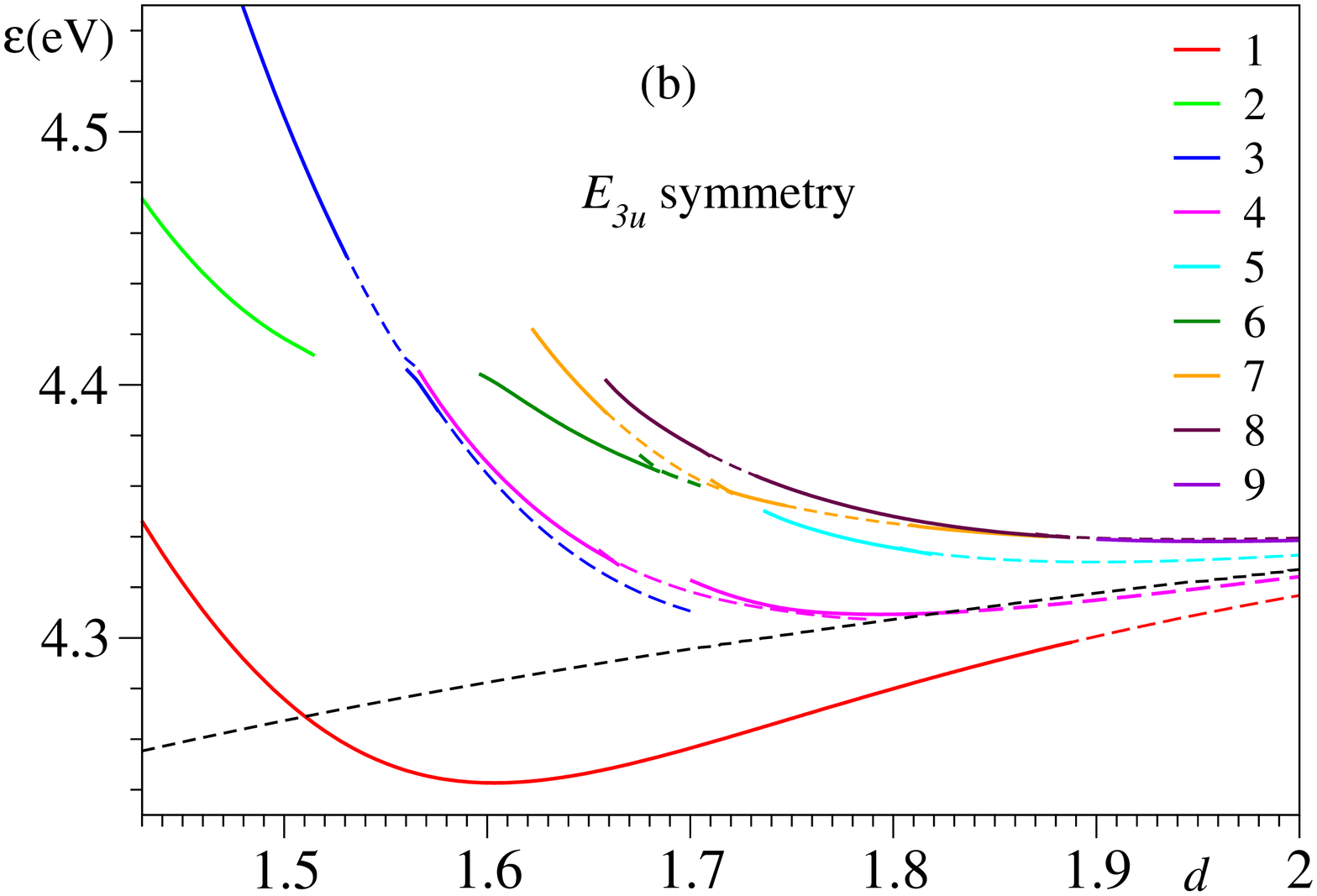}
}
\centerline{
\includegraphics[width=0.35\textwidth,angle=-90]{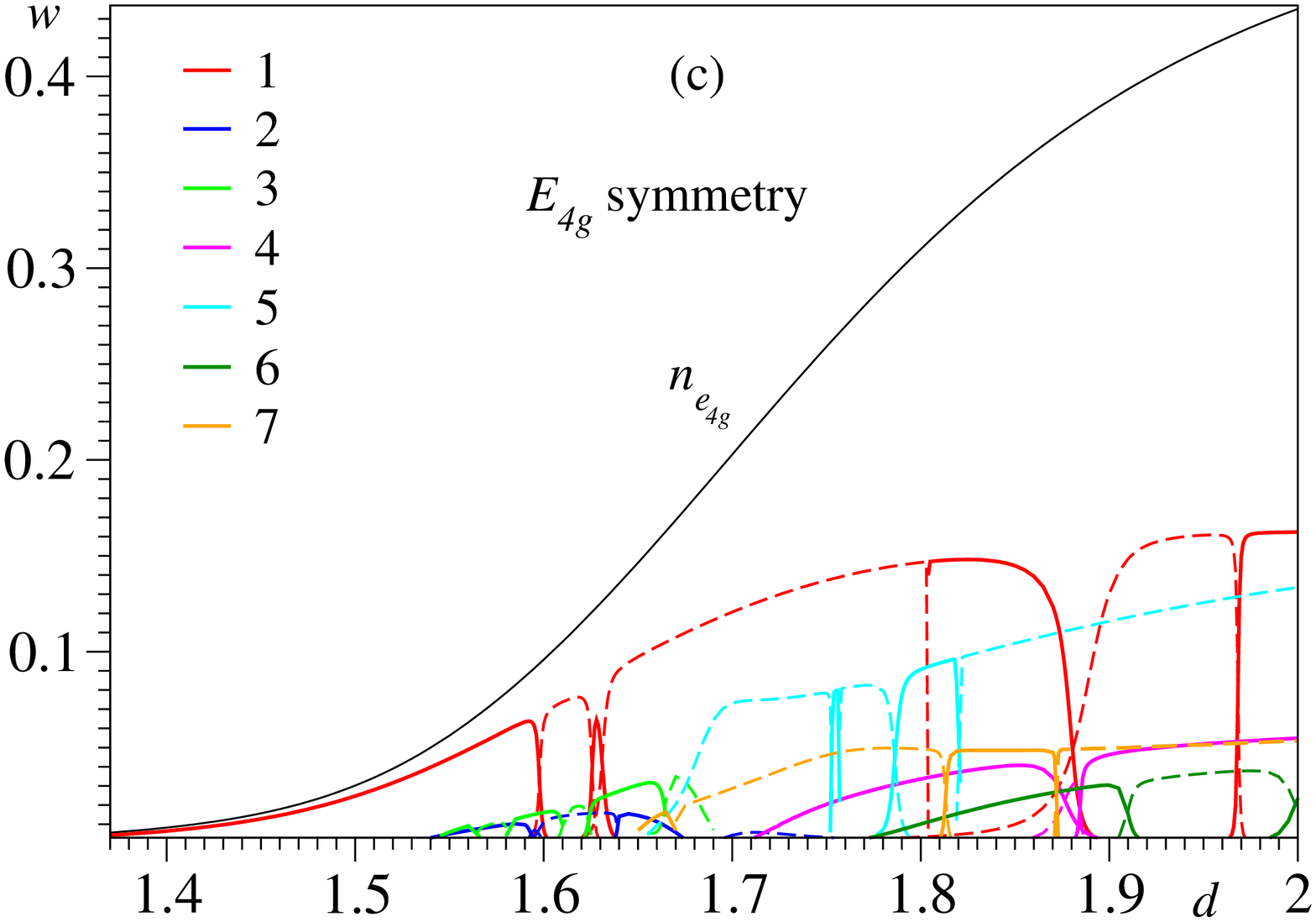}
\includegraphics[width=0.35\textwidth,angle=-90]{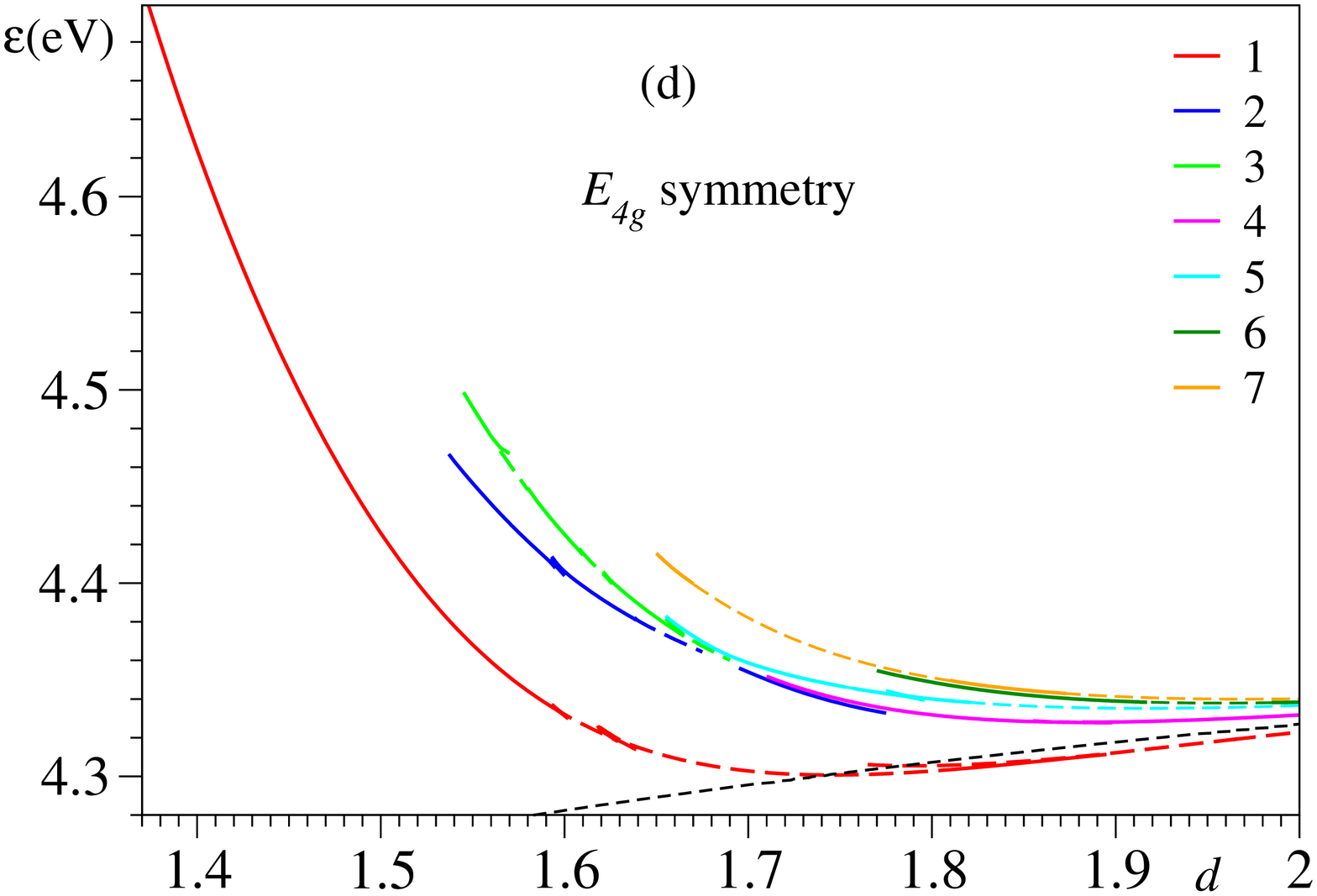}
}
\centerline{
\includegraphics[width=0.35\textwidth,angle=-90]{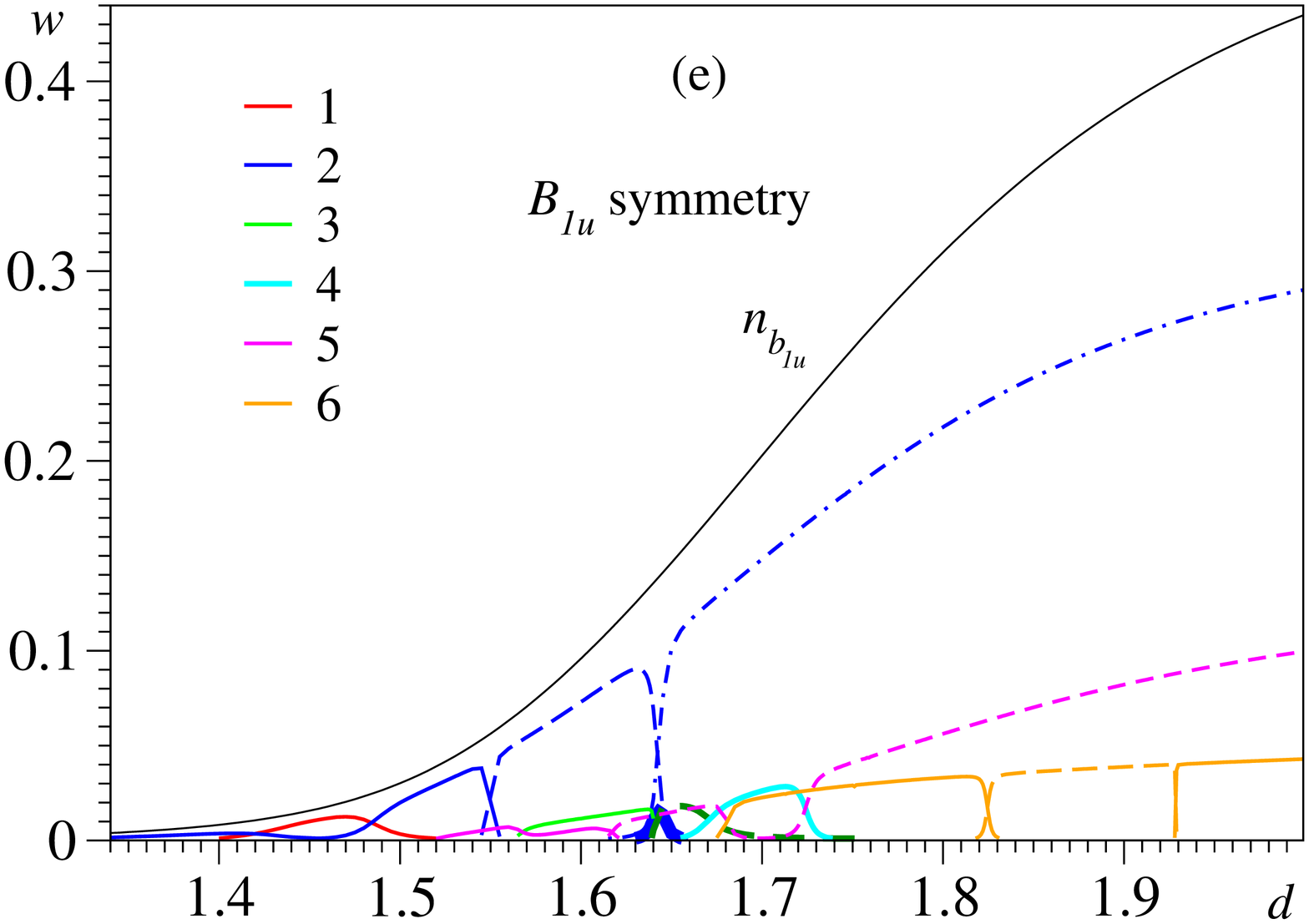}
\includegraphics[width=0.35\textwidth,angle=-90]{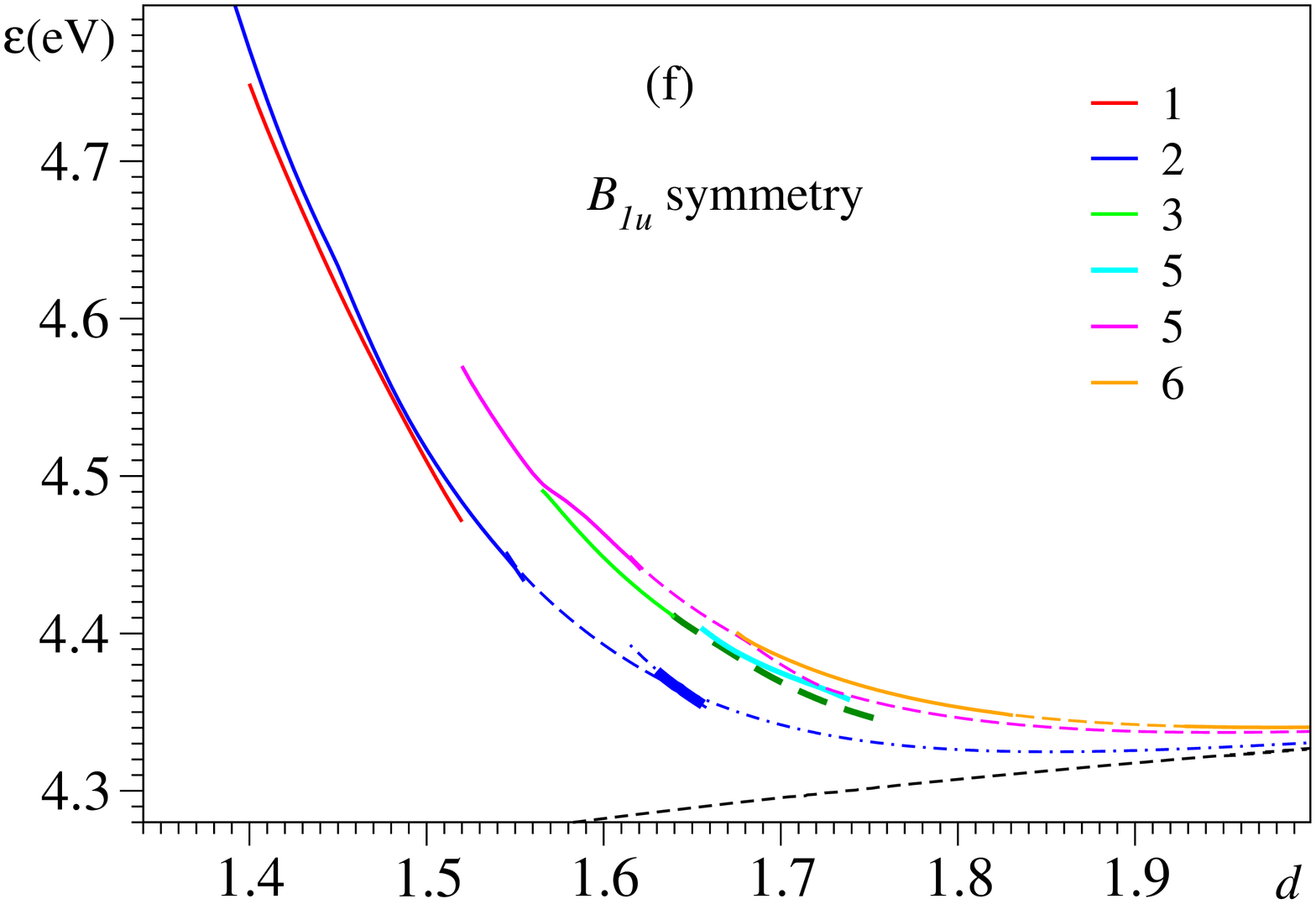}
}
\caption{(Color online) $E_{3u}$--, $E_{4g}$--, and $B_{1u}$--spectral factors $w$
and ionization energies $\varepsilon$ versus interdot spacing $d$ in    
ten--QD nanorings. The numbers $i=1,2,\ldots$ in the legend label the ionized eigenstates  
$\Psi_{k,i}$. The black dashed line corresponds to the lowest 
ionization energy $ - \varepsilon_{H} - 2 V$ in the limit of perfect localization.}
\label{fig:E3-E4-B1-10-QDs}
\end{figure}

The numerical results for ionization in ten--dot nanorings are presented in 
Figs.\ \ref{fig:A1-E1-E2-10-QDs} and \ref{fig:E3-E4-B1-10-QDs},
separately for each symmetry. They confirm the general features 
already observed in six--QD nanorings. 
For the ionization processes having the symmetries ($A_{1g}$, $E_{1u}$, and $E_{2g}$) 
of the MOs completely occupied within the MO--picture (let us call them ``occupied'' MOs), 
there exist one signal (let us call it the ``main'' signal) per each symmetry whose 
spectroscopic factor is close to the ideal value $w_{MO}=1$ at $d\agt 1$.
With increasing $d$, it decreases from $w_{MO}=1$ 
and becomes more and more smaller than the corresponding MO--population 
($n_0\equiv n_{A_{1g}}$, $n_1\equiv n_{E_{1u}}$, $n_2\equiv n_{E_{2g}}$). 
Concomitantly, a few more (``satellite'') signals aquire more and more 
significant spectral weight, such that at larger $d$ 
they can even dominate over the ``main'' signal. 
See Fig.\ \ref{fig:A1-E1-E2-10-QDs}. Interestingly, 
the deeper the ``occupied'' MO, the more is it affected 
by correlations at larger $d$; the relative weight 
of the ``satellite'' signals to the ``main'' signal for $A_{1g}$--symmetry 
(deepest ``occupied'' MO) is larger than that for  $E_{1u}$--symmetry, which is in turn larger 
than for  $E_{2g}$--symmetry (highest ``occupied'' molecular orbital, HOMO). 
\par
At very small interdot distances ($d\agt 1$), the ionization from the ``unoccupied'' MOs 
(\emph{i.\ e.}, those unoccupied if the MO--picture were valid) is altogether ineffective.
However, as a result of the strong electron correlations, the spectroscopic factors of these 
$E_{3u}$--, $E_{4g}$--, and $B_{1u}$--ionization processes gradually 
increase with increasing $d$. Their spectral weight also becomes distributed 
over several lines. Similar to that from the ``occupied'' MOs, these 
spectroscopic factors are comparable among themselves, and also with those of the relevant 
ionization signals of the ``occupied'' MOs.
See Figs.\ \ref{fig:E3-E4-B1-10-QDs} and compare them 
with Figs.\ \ref{fig:A1-E1-E2-10-QDs}. 
\par
To conclude, 
similar to the case of six--dot nanorings of Sec.\ \ref{sec:6-QDs}, the MO--picture 
rapidly deteriorates with increasing $d$. 
\par
From the inspection of the ionization factors of six-dot nanorings 
(Figs.\ \ref{fig:A1-E1-6-QDs}, \ref{fig:E2-6-QDs}, and \ref{fig:B1-6-QDs}), 
one may be tempted to think that avoided crossings only occur for 
two--dimensional irreducible representations
($E_{1u}$ and $E_{2g}$ in that case) and not for the one-dimensional irreducible representations 
($A_{1g}$ and $B_{1u}$). However, this is not true, as illustrated by the results 
for the one--dimensional irreducible representations of ten--dot nanorings 
presented in Figs.\ \ref{fig:A1-E1-E2-10-QDs}a and \ref{fig:E3-E4-B1-10-QDs}e. 
For $A_{1g}$--symmetry, two avoided crossings are visible in 
Figs.\ \ref{fig:A1-E1-E2-10-QDs}a and b: 
one involving the states denoted by 3 and 4 at $d\simeq 1.561$, another 
with the participation of the states 4 and 5 at  $d\simeq1.670$. 
The corresponding minimum energy differences 
are $5.33$\,meV and $0.26$\,meV, respectively. As seen in Figs.\ \ref{fig:E3-E4-B1-10-QDs}e and f, 
the avoided crossings for $B_{1u}$--symmetry are more numerous. 
One of them, that at $d\simeq 1.643$, even involves three states.

\par
Similar to the case of six--dot nanorings (Sec.\ \ref{sec:6-QDs}), 
one can compare the information deduced by ionization for ionized nanorings 
with that obtained by optical absorption for nanorings with nine electrons and ten QDs.
As discussed in Ref.\ \onlinecite{Baldea:2008}, in the latter case
one can target eigenstates of $E_{1u}$-- and $E_{3u}$--symmetry.
\par
Certain avoided crossings are visible both in ionization and in optical absorption spectra.
This is the case for the avoided crossing at $d\simeq 1.755$ involving the states 2 and 3, as well as the 
pretty broad one around $d\approx 1.87$.
On the contrary, 
the dense avoided crossings occuring in the narrow range $1.723 \alt d\alt 1.737$, 
which involve the states 9, 10, 11, and 12 can be seen only in ionization 
(Fig.\ \ref{fig:A1-E1-E2-10-QDs}c and d),
but not in optical absorption, because of the reduced intensities in the latter case; see Fig.\ 9d 
of Ref.\ \onlinecite{Baldea:2008}.
\par
The fact that more than two states can participate to an avoided crossing was already noted 
in the study on optical absorption \cite{Baldea:2008}. An interesting situation can be seen 
in the  $E_{3u}$--spectrum (Fig.\ \ref{fig:E3-E4-B1-10-QDs}a and b), 
where two avoided crossings are visible 
around $d\simeq 1.885$: one at lower energies involves two states, another at higher energies 
with the participation of three states. These avoided crossings are insignificant for optical 
absorption, because of the reduced spectral intensity of these states \cite{Baldea:2008}. 
\par
A somewhat reversed situation occurs in the $E_{3u}$--optical spectrum,  
where the participation of three states to the avoided crossing at $d\simeq 1.525$  
(see Fig.\ 9f of Ref.\ \onlinecite{Baldea:2008}), while only two are significant 
(visible) in the $E_{3u}$--ionization spectrum of Fig.\ 
\ref{fig:E3-E4-B1-10-QDs}a.
\section{Avoided Crossings}
\label{sec:avoided-crossings}
\begin{figure}[h]
\centerline{\hspace*{-0ex}
\includegraphics[width=0.35\textwidth,angle=-90]{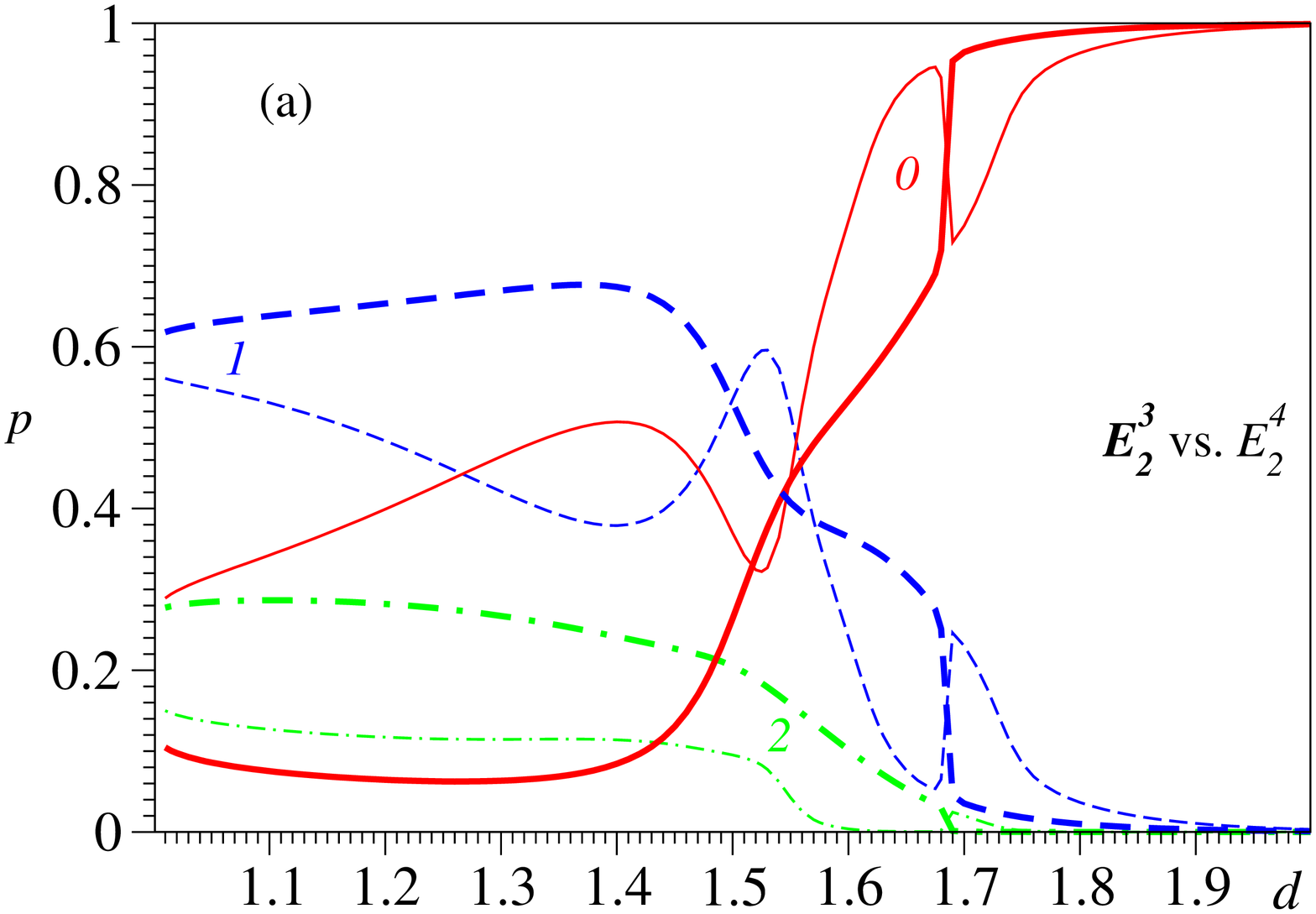}
\includegraphics[width=0.35\textwidth,angle=-90]{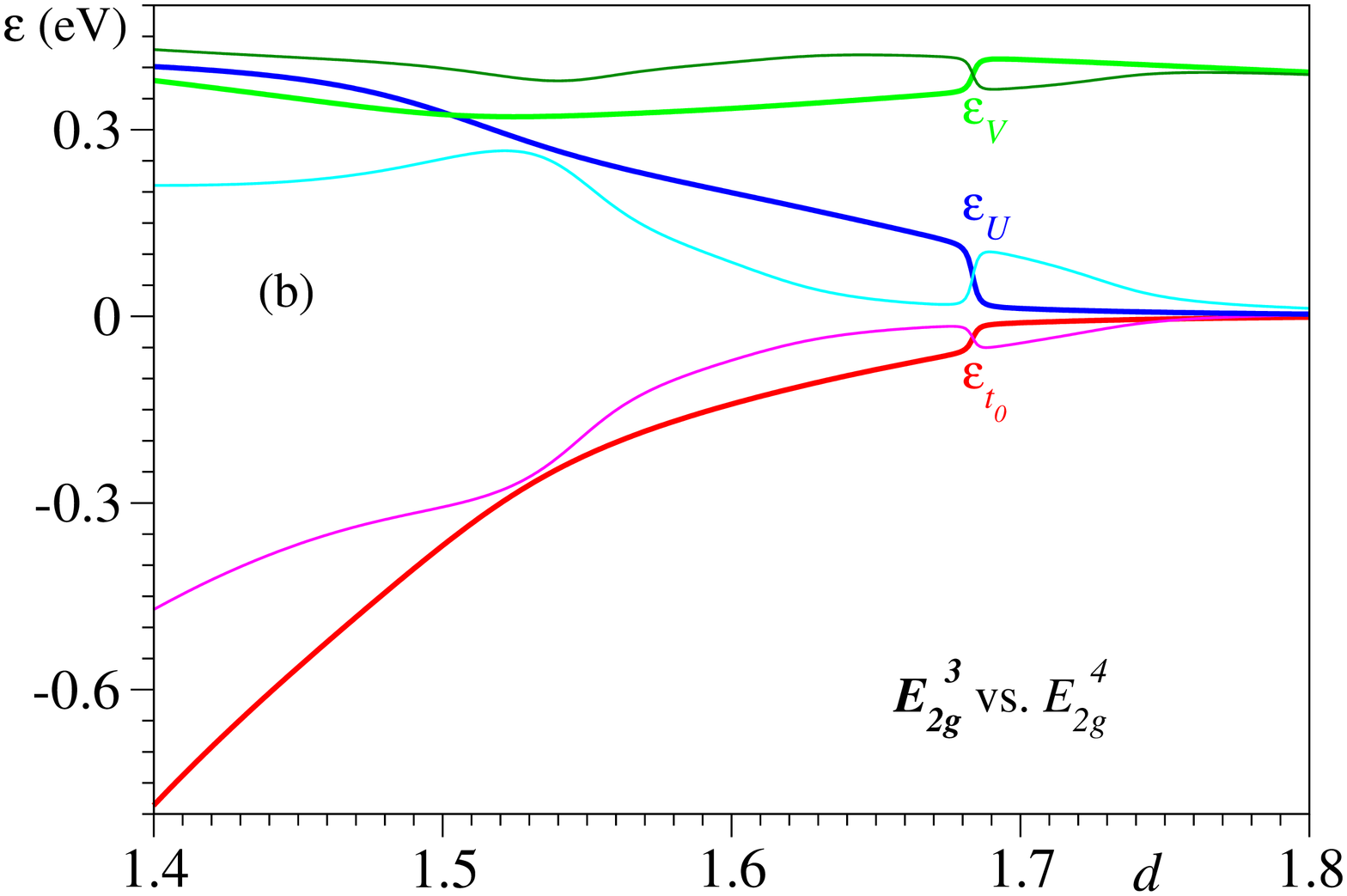}
}
\caption{(Color online) $d$-dependence of the weights $p_0, p_1, p_2$ 
of the various multielectronic configurations (a) 
and the separate contributions of the $t_0$--, $U$--, and $V$-- terms 
to the total energy (b) of the ionized eigenstates 
$E_{2g}^{3}$ (thick lines) vs.\ $E_{2g}^{4}$ (thin lines). See the main text and the caption of Fig.\ 
\ref{fig:S}.}
\label{fig:5el-E2_3-E2_4-QDs}
\end{figure}

The fact that avoided crossings involving energy curves of eigenstates of identical symmetries 
represent a frequent phenomenon in metallic QD--nanorings has already noted 
in the study of optical absorption \cite{Baldea:2008}. 
As a general characterization, as visible in Figs.\ 
\ref{fig:E2-6-QDs}a, \ref{fig:A1-E1-E2-10-QDs}a, \ref{fig:A1-E1-E2-10-QDs}c, \ref{fig:A1-E1-E2-10-QDs}e, 
\ref{fig:E3-E4-B1-10-QDs}a, \ref{fig:E3-E4-B1-10-QDs}c, and \ref{fig:E3-E4-B1-10-QDs}e 
they only occur at larger $d$, where correlations are important. 
Two (or more) states of identical symmetry 
described by a single Slater determinant, where the MOs are either occupied or empty,  
cannot come too close in energy; basically, their energy difference is determined by the energy differences 
of their occupied MOs. Things change at larger $d$, where $t_0$ ceases to be 
the dominant energy scale and frustration due to the $U$-- and $V$--terms 
becomes important. For illustration, we present in Fig.\ \ref{fig:5el-E2_3-E2_4-QDs} results 
for the pair of $E_{2g}$-states ($E_{2g}^3$ and $E_{2g}^4$) 
of the six-QD nanoring (\emph{cf.\ }Fig.\ \ref{fig:E2-6-QDs}).
As seen Fig.\ \ref{fig:5el-E2_3-E2_4-QDs}a, at the point $d \simeq 1.685$ 
the weights of the multielectronic configurations where none, one, or 
two QDs are doubly occupied rapidly interchange between the mates involved 
in the avoided crossing. The contributions to energy of the 
$t_0$--, $U$--, and $V$--terms, \emph{i.\ e.}, 
$-t_0 \sum_{l,\sigma}  
\langle \Psi_{k,i}\vert (a_{l,\sigma}^{\dagger} a_{l+1,\sigma}^{} + \mbox{h.c.})\vert \Psi_{k,i}\rangle
$,
$U \sum_{l}  
\langle \Psi_{k,i}\vert  n_{l,\uparrow}^{} n_{l,\downarrow}^{} \vert \Psi_{k,i}\rangle
$, and  
$V \sum_{l}  
\langle \Psi_{k,i}\vert  n_{l} n_{l+1} \vert \Psi_{k,i}\rangle
$, respectively are sensitive to dot occupations. Indeed, their curves at the avoided crossing, 
depicted in Fig.\ \ref{fig:5el-E2_3-E2_4-QDs}b,  
behave accordingly. 
\par
\begin{figure}[h]
\centerline{\hspace*{-0ex}
\includegraphics[width=0.35\textwidth,angle=-90]{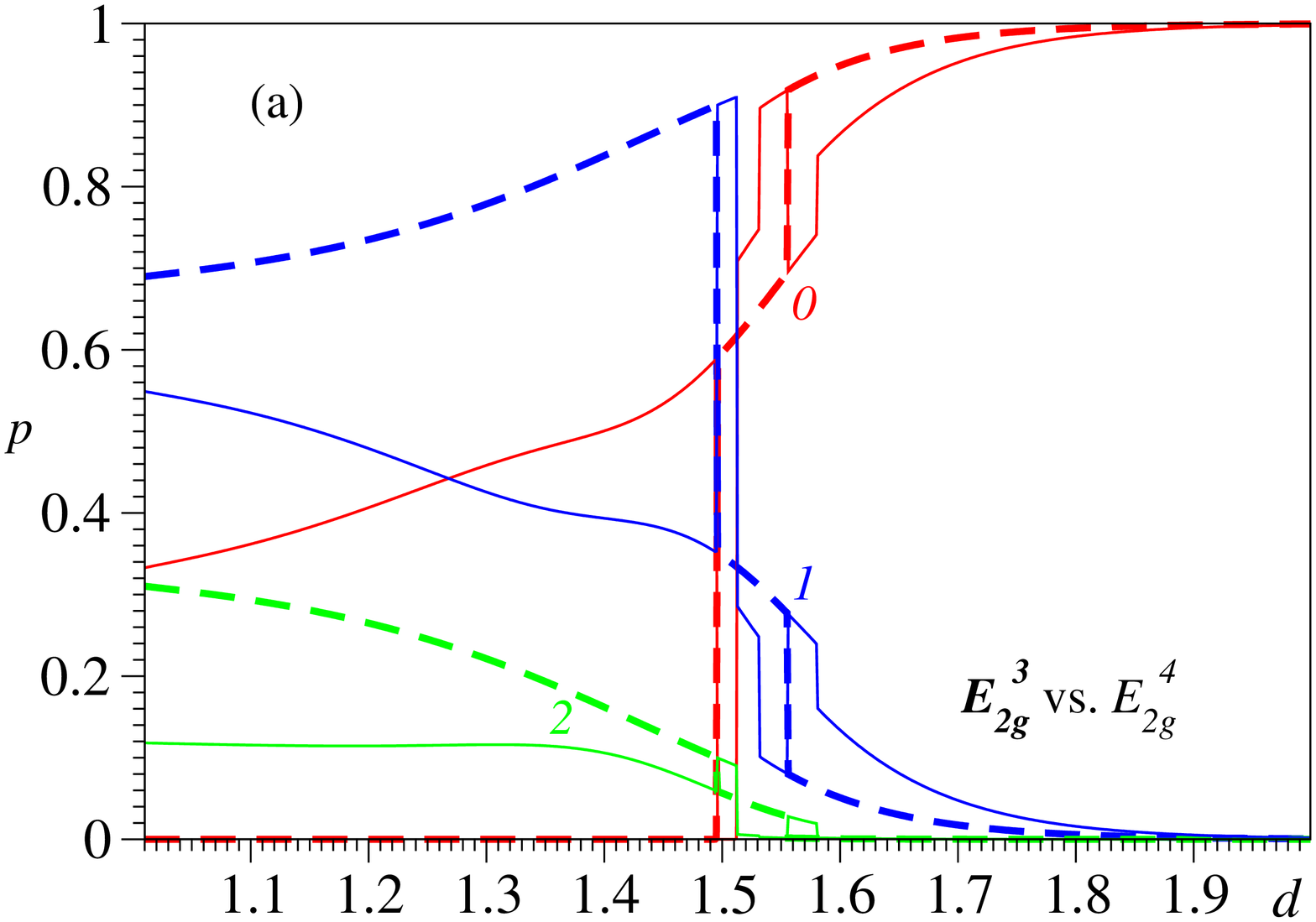}
\includegraphics[width=0.35\textwidth,angle=-90]{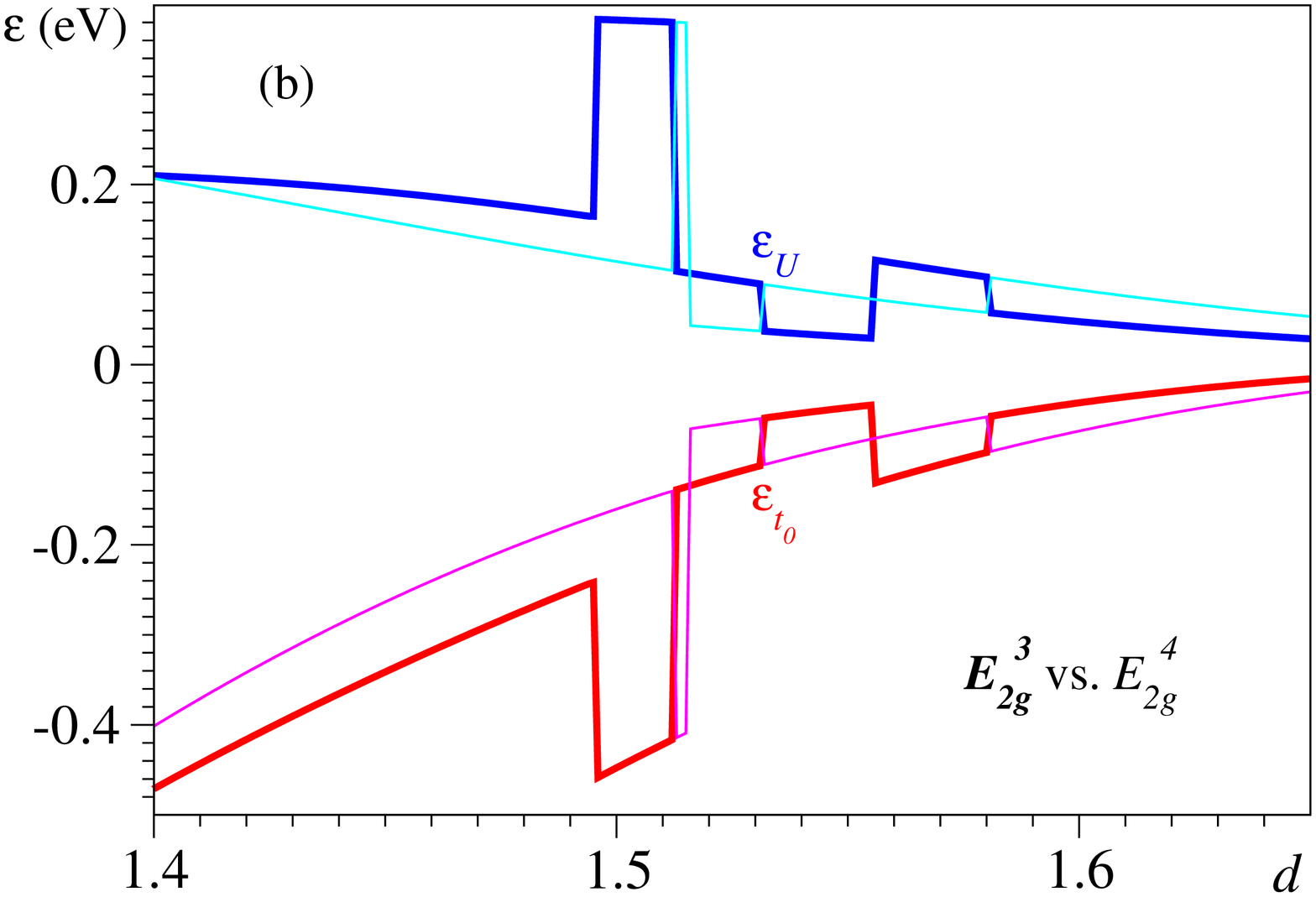}
}
\caption{(Color online) $d$-dependence of the weights $p_0, p_1,p_2$ 
of the various multielectronic configurations (a) 
and separate contributions of $t_0$-- and $V$-- terms 
to the total energy (b) of the ionized eigenstates 
$E_{2g}^{3}$ (thick lines) vs.\ $E_{2g}^{4}$ (thin lines) in the case $V=0$. See the main text and the caption of Fig.\ 
\ref{fig:S}.}
\label{fig:5el-E2_3-E2_4-QDs-V=0}
\end{figure}

One may ask at this point whether 
avoided crossings only occur in the \emph{extended} Hubbard ring, or also 
in the \emph{restricted} (\emph{i.\ e.}, $V=0$) Hubbard ring. 
The results presented in 
Fig.\ \ref{fig:5el-E2_3-E2_4-QDs-V=0}, which represents the counterpart of 
Fig.\ \ref{fig:5el-E2_3-E2_4-QDs} for $V=0$,
reveal that they are also present in the latter case. 
Again, the physics behind avoided crossings is the interplay between 
the competing terms (in this case $t_0$- and $U$-terms). 
For each eigenstate involved in an avoiding crossing, the 
individual ($t_0$-- and $U$--)terms exhibit 
jumps at the avoided crossing point that compensate each other in their sum (the total energy), 
which is represented by a smooth curve around this point. 
To avoid confusions, we note that the jumps at $d\simeq 1.495$ visible in 
Fig.\ \ref{fig:5el-E2_3-E2_4-QDs-V=0} for the curves of the state $E_{2g}^{3}$ 
are related to another avoided crossing between the states $E_{2g}^{3}$ and $E_{2g}^{2}$.
The latter, which is not shown in the figure, is characterized by jumps opposite 
to those for the  $E_{2g}^{3}$--curve.
\par
From a pragmatic standpoint, 
one may wonder whether avoided crossings are of importance at all; since what one 
can measure there is merely a smooth bright diabatic state, and not two 
(or several) individual (adiabatic) states with identical symmetry 
exhibiting rapid variations. 
However, one should mention that this applies to the case of the 
\emph{ideal} nanoring with equidistant QDs. In a distorted nanoring, 
there will be avoided crossings between the ionization energies of 
the different symmetries of the ideal nanoring. 
Besides, similar to ordinary molecules \cite{koeppel:84}, strong nonadiabatic effects 
can be expected in the presence of phonons.
\section{Hidden Quasi--Symmetry}
\label{sec:hidden-symmetry}
Most importantly, this investigation gives further support to the  
most intriguing aspect of electron correlations previously encountered 
in optical absorption \cite{Baldea:2008}:
in spite of the fact that correlations are strong, out of very numerous 
ionized eigenstates allowed by spatial symmetry to contribute to the ionized spectrum, 
only very few states possess a significant spectroscopic factor.
\par
The ionization spectra depicted in Figs.\ \ref{fig:full-spectra}a and b 
are indeed richer than those, which comprise two and three lines, respectively, 
obtained within the MO approach. However, the number of the ionization signals 
is much smaller than the number of states, which are expected to contribute in a strongly correlated system.
To properly assess the intriguing aspect of correlations, one should compare 
the number of significant ionization signals with the substantially larger 
number of states allowed to contribute by spatial symmetry and spin conservation.
For the six--dot nanoring of Figs.\ \ref{fig:full-spectra}a, 
there are ten significant ionization signals but 178 allowed transitions.   
For the ten--dot nanorings, the number of $15$ lines visible in 
Figs.\ \ref{fig:full-spectra}b should be compared with the number $\sim 10^4$ of states allowed 
by spatial symmetry and spin conservation.
\par
\begin{figure}[h]
\centerline{
\includegraphics[width=0.35\textwidth,angle=-90]{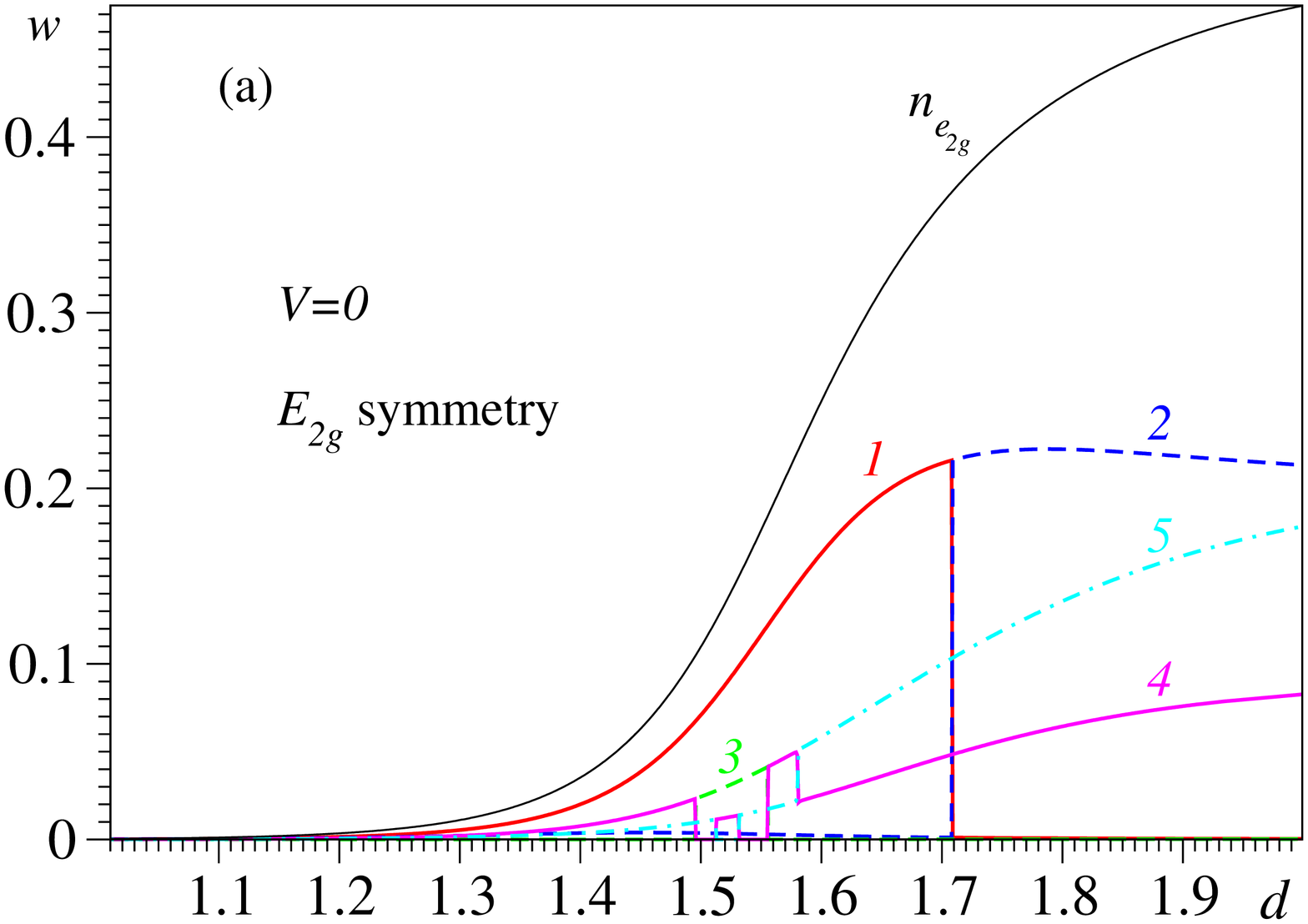}
\includegraphics[width=0.35\textwidth,angle=-90]{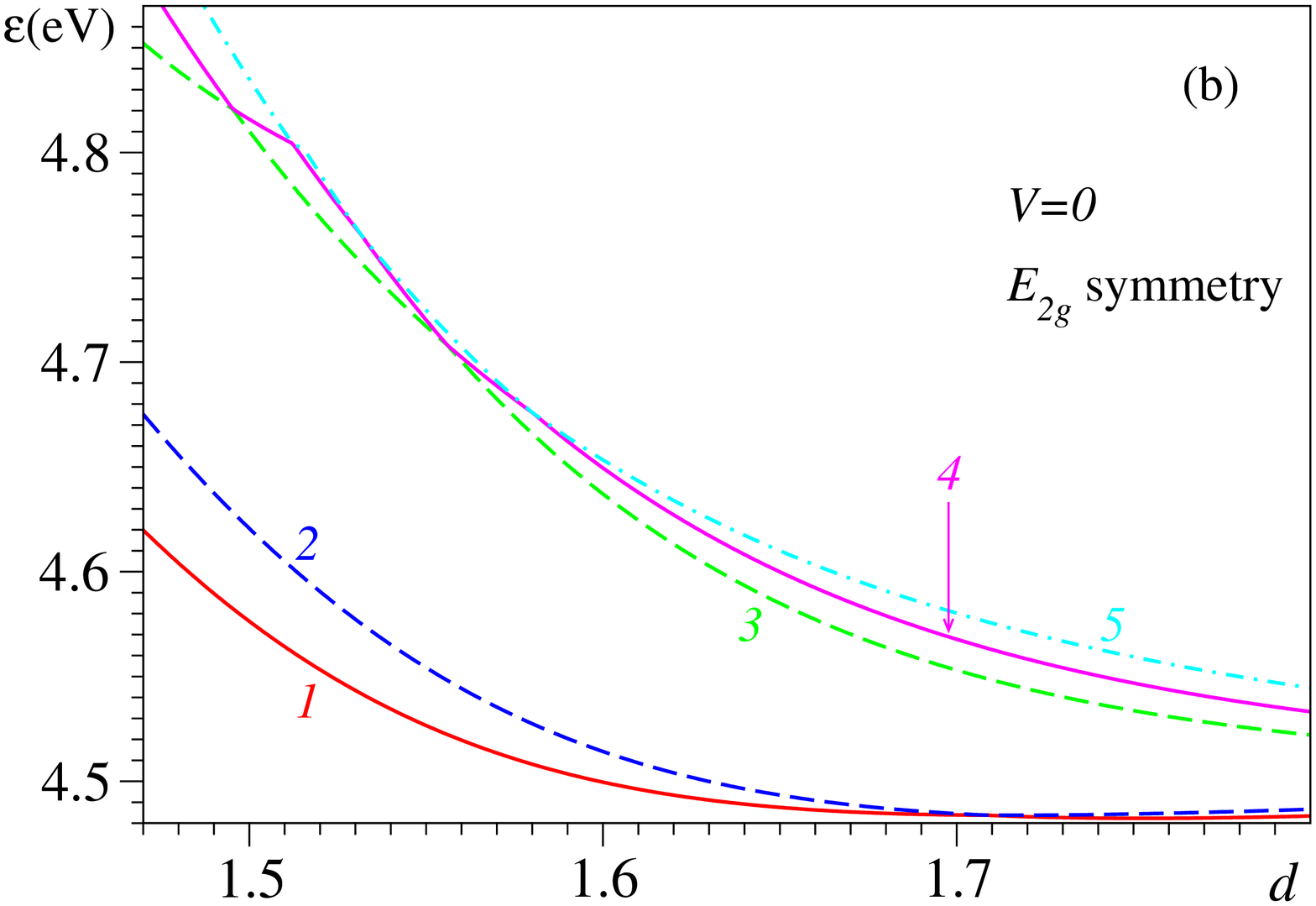}
}
\centerline{\hspace*{-0ex}
\includegraphics[width=0.35\textwidth,angle=-90]{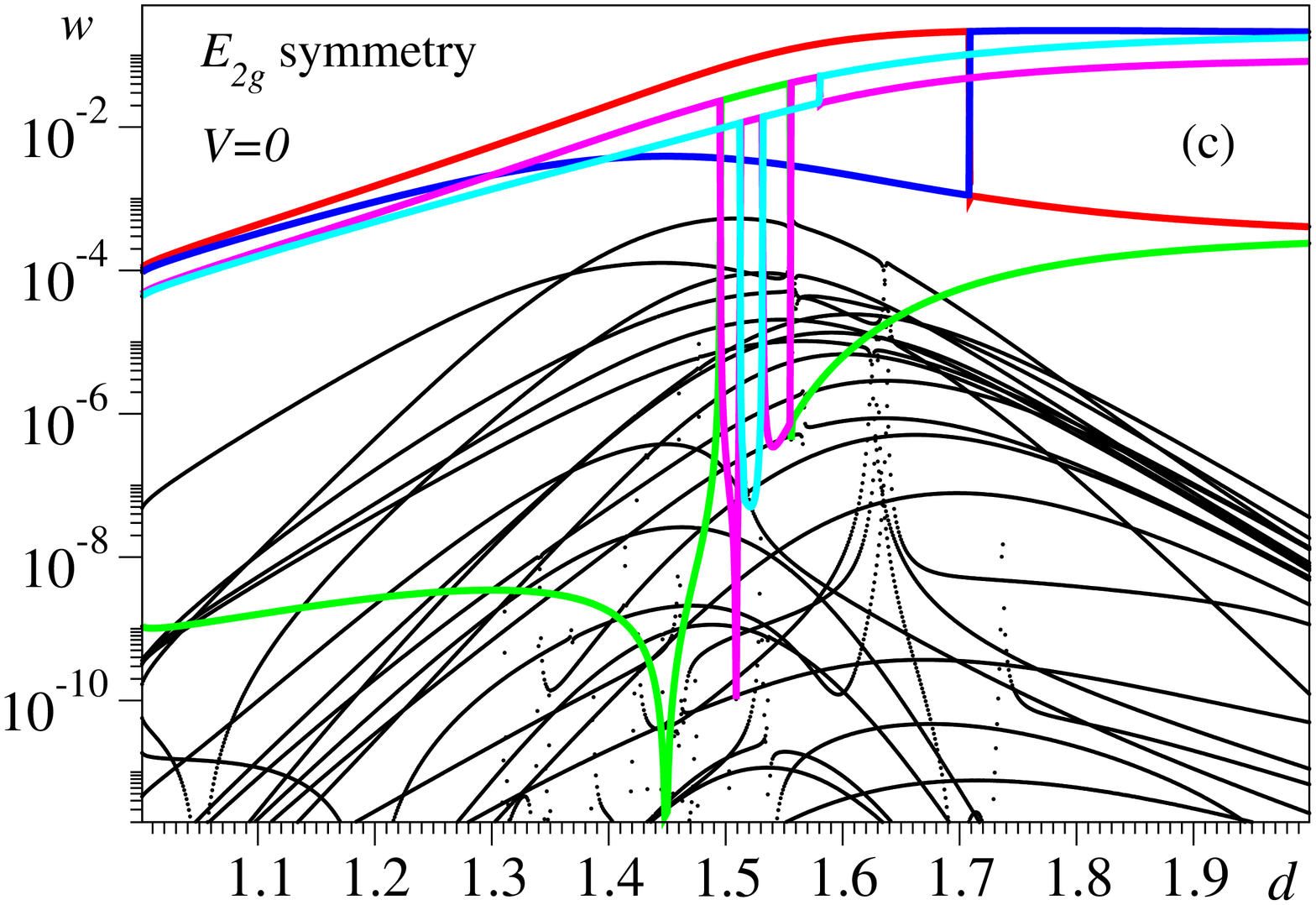}
}
\caption{(Color online) $E_{2g}$--spectral factors $w$ (a) and ionization energies $\varepsilon$ (b)
versus interdot spacing $d$ in    
six--QD nanorings for $V=0$ (counterpart of Fig.\ \ref{fig:E2-6-QDs}). 
The numbers $i=1,2,\ldots$ in the legend 
label the ionized eigenstates $\Psi_{k,i}$ [\emph{cf.\ }Eq.\ (\ref{eq-w})].
Notice the numerous spectral lines with very small intensities in panel (c), 
which represents panel (a) redrawn using 
the logarithmic scale on the ordinate, demonstrating that the restricted Hubbard model 
is also characterized by a hidden \emph{quasi}--symmetry.}
\label{fig:E2-6-QDs-V=0}
\end{figure}

In addition to well established 
symmetries \cite{andrei,andrei-preprint,LiebWu:03,Shastry:86,Shastry:88,Grosse:89,Wadati:87,Olmedilla:88},
the restricted (\emph{i.\ e.}, $V=0$) Hubbard model for rings is known to possess 
hidden symmetries \cite{RamosHiddenSymmetries:97}. 
Similar to the case of optical absorption, the scarcity of the ionized 
spectra of nanorings described by the extended Hubbard model discussed here 
points towards a hidden \emph{quasi}--symmetry: besides a few signals with significant 
spectroscopic factors, there are numerous lines of very small but definitely non-vanishing 
intensities. These almost vanishing signals, which exhibit a regular $d$--dependence, 
can only be seen in Fig.\ \ref{fig:E2-6-QDs}c, 
due to the logarithmic scale on the ordinate. In all of the other figures presented so far,
they are invisible within the drawing accuracy.
Therefore, one may be tempted to think that this hidden quasi-symmetry will 
evolve into a true hidden symmetry as $V\to 0$. However, this is not the case,
and for illustration we present in Fig.\ \ref{fig:E2-6-QDs-V=0} results 
for $E_{2g}$--ionization in six--QD nanorings for $V=0$, 
which represents the counterpart of Fig.\ \ref{fig:E2-6-QDs} at $V\neq 0$.
The presence in Fig.\ \ref{fig:E2-6-QDs-V=0}c 
of numerous weak but non--vanishing spectral factors, also displaying a 
regular $d$--dependence and being  
many magnitude orders larger than inherent numerical inaccuracies, 
demonstrates that the restricted Hubbard model 
is also characterized by a hidden \emph{quasi}--symmetry, similar to the extended Hubbard model.
\par
Of course, the actual task of theory 
is to specify the hidden quasi-symmetry more precisely.
An interesting observation concerns the number of significant spectral lines per each symmetry.
As already noted, the number $N_i$ of significant ionization signals is larger  
than expected within the MO picture, or, rephrasing, larger
than the number $N_{1h}$ of one-hole processes possible in the neutral 
self--consistent--field (SCF)-ground state.
However, this number can be comprehended if one considers in addition to $N_{1h}$
the number $N_{2h-1p}$ of two-hole--one-particle 
processes possible in the neutral SCF-ground state, wherein the excited electron 
occupies the MO just above the Fermi level. That is, the observation is  
that $N_i = N_{1h} + N_{2h-1p}$.
\par
For instance, for a six-QD nanoring, there are two $1h$-processes, 
one with $A_{1g}$-- and another with $E_{1u}$--symmetry. 
The aforementioned $2h$-$1p$-processes are represented by the products 
$E_{1u}^{-1} \times E_{1u}^{-1} \times E_{2g}$,
$E_{1u}^{-1} \times A_{1g}^{-1} \times E_{2g}$, 
$A_{1g}^{-1} \times E_{1u}^{-1} \times E_{2g}$, and 
$A_{1g}^{-1} \times A_{1g}^{-1} \times E_{2g}$. 
By using the multiplication rules of D$_{6h}$, they lead to 
one $A_{1g}$-, two $E_{1u}$-, four $E_{2g}$-, and two $B_{1u}$-processes. 
In this way, we arrive at two $A_{1g}$-, three $E_{1u}$-, four $E_{2g}$-, 
and two $B_{1u}$-ionization processes. This is just what one observes in 
Figs.\ \ref{fig:A1-E1-6-QDs}a, \ref{fig:A1-E1-6-QDs}c, \ref{fig:E2-6-QDs}, and 
\ref{fig:B1-6-QDs}, respectively, if one considers the diabatic bright state at 
avoided crossings \cite{avoided-crossings}. 
\par
For ten--QD nanorings, there are three $1h$--processes, 
of $A_{1g}$--, $E_{1u}$--, and $E_{2g}$--symmetry.
The analysis is similar but more tedious for the relevant $2h$-$1p$-processes. 
They are represented by the products
$E_{2g}^{-1} \times E_{2g}^{-1} \times E_{3u}$,
$E_{2g}^{-1} \times E_{1u}^{-1} \times E_{3u}$,
$E_{2g}^{-1} \times A_{1g}^{-1} \times E_{3u}$,
$E_{1u}^{-1} \times E_{2g}^{-1} \times E_{3u}$,
$E_{1u}^{-1} \times E_{1u}^{-1} \times E_{3u}$,
$E_{1u}^{-1} \times A_{1g}^{-1} \times E_{3u}$,
$A_{1g}^{-1} \times E_{2g}^{-1} \times E_{3u}$,
$A_{1g}^{-1} \times E_{1u}^{-1} \times E_{3u}$, and 
$A_{1g}^{-1} \times A_{1g}^{-1} \times E_{3u}$.
In view of the multiplication table of the point group D$_{10,h}$,
one gets three $A_{1g}$-, five $E_{1u}$--, five $E_{2g}$--, six $E_{3u}$--, 
six $E_{4g}$--, and three $B_{1u}$--processes altogether. Again, these numbers agree with 
the numbers of significant ionization signals visible in Figs.\ 
\ref{fig:A1-E1-E2-10-QDs} and \ref{fig:E3-E4-B1-10-QDs}, respectively.
\par
The fact that for six--QD nanorings we can compute exactly \emph{all} eigenstates enables 
us to continuously vary the model parameters and find out the counterparts 
of the states with significant spectroscopic factors in the 
limit $d \agt 1$, where $U, V \ll 4 t_0$ and the MO--picture is reliable.  
The energies of the aforementioned 
$1h$-- and $2h$-$1p$--processes are given in Table \ref{tab:1} for the limit $U, V \to 0$.
For $d \to 1$, one finds $t_0 \simeq 1.49$\,eV , $U/4 t_0 \simeq 0.049$, and $V/4t_0 \simeq 0.036$. 
By inspecting the curves for the ionization energies of Figs.\ \ref{fig:A1-E1-6-QDs}b, 
\ref{fig:A1-E1-6-QDs}d, \ref{fig:E2-6-QDs}, and \ref{fig:B1-6-QDs} one can see that they 
compare favorably to those of Table \ref{tab:1}. In this limit, the main effect 
of interaction is to split the degenerate energies by amounts of the 
order of $U$ and $V$. This demonstrates that, indeed, the ionization processes 
with significant spectroscopic factors evolve into the 
$1h$-- and $2h$-$1p$--processes occurring in the uncorrelated case.
\begin{table}
\caption{The energies of the $1h$-- and $2h$-$1p$--processes described in the main text 
for six--QD nanorings 
in the limit $U, V \to 0$.}
\label{tab:1}
\begin{tabular}{cccc}
\hline
Symmetry & \hspace*{4ex} Process \hspace*{4ex} & 
\hspace*{4ex} Ionization Energy \hspace*{4ex} & Degeneracy\\
\hline
$E_{1u}$  & $1h$     & $ -\varepsilon_H +  t_0 $ & 1  \\
         & $2h$-$1p$ & $ -\varepsilon_H + 4 t_0 $ & 2  \\
\hline
$A_{1g}$  & $1h$      & $ -\varepsilon_H +  2 t_0 $ & 1  \\
         & $2h$-$1p$ & $ -\varepsilon_H + 3 t_0 $ & 1  \\
\hline
$E_{2g}$  & $2h$-$1p$ & $ -\varepsilon_H +  3 t_0 $ & 3  \\
         & $2h$-$1p$ & $ -\varepsilon_H + 5 t_0 $ & 1  \\
\hline
$B_{1u}$  & $2h$-$1p$ & $ -\varepsilon_H +  4 t_0 $ & 2  \\
\hline
\end{tabular}
\end{table} 
\par
For a quantitative description suggested by the above considerations, 
it appears most straightforwardly 
to employ $c_{k,\sigma}\vert \Phi\rangle$ and 
$c_{k,\sigma}c_{k_1,\sigma_1}^{\dagger} c_{k_2,\sigma_2}\vert \Phi\rangle$
with $\vert k\vert, \vert k_2\vert \leq k_F$ and $\vert k_1\vert =k_F + 1$
for constructing linear independent vectors with appropriate spin and spatial symmetries.
Here, $k_F$ denotes the Fermi wave vector 
($k_F=n$ for closed shell rings with $N=4n+2$, $n$ being an integer)
and $\vert \Phi\rangle$ the exact neutral ground state \cite{scf-poor}.
We have used them as basis (sub)set for diagonalization,  
and found approximate eigenvectors $\tilde{\Psi}_{k,i}$ of the ionized nanoring. 
They have been utilized to compute an approximate ionization spectrum via Eq.\ (\ref{eq-w}). 
Apart from the fact that the number of lines is correctly obtained, 
this straightforward approach fails, however, to quantitatively reproduce the ionization spectra. 
\par
In the subsequent attempt to reproduce the ionization spectra,
we have generalized this approach, by lifting the above constraints 
imposed on $k, k_1$, and $k_2$. In a further effort, we have alternatively employed the 
exact ionized ground state $\Psi$ and the dressed particle--hole 
excitations $c_{k_1,\sigma_1}^{\dagger} c_{k_2,\sigma_2}\vert \Psi\rangle$
to construct a basis subset for diagonalization. 
Both aforementioned approximations are in the spirit of the 
treatment based on dressed particle-hole excitations 
developed in Ref.\ \onlinecite{Baldea:2008}, where it turned out to be 
an insightful approximate method to study optical absorption.
Unfortunately, none of these two methods is able to quantitatively reproduce 
the exact ionization spectra of strongly correlated QD--nanorings.   
\par
Similar to the case of optical absorption \cite{Baldea:2007,Baldea:2008},
in spite of a variety of attempts of analyzing the numerical results,
unfortunately, we cannot offer a quantitative explanation of the hidden quasi--symmetry 
behind the scarcity of the ionization spectrum of strongly correlated nanorings. 
What we can do is only to tentatively speculate on the nature of this symmetry.
Namely, based on the above considerations, we claim 
the existence of a one-to-one correspondence between 
the $1h$-- and $2h$-$1p$--processes possible in the SCF--neutral ground state 
and the number of significant lines in the ionization spectrum of 
strongly correlated half--filled 
QD-nanorings described by the extended Hubbard model.
The basic postulate of the Landau theory \cite{Landau:1956,Landau:1957,Landau:1959} 
is the one--to--one map between the low-energy excitations of noninteracting and 
interacting electron systems. 
Applied to ionization, Landau theory predicts a one--to--one correspondence 
between the ionization signals and the one--hole ($1h$) ionized states, a fact 
contradicted by the exact ionization spectra.  
Therefore, our hypothesis on the number of relevant lines represents an 
extension of Landau's basic idea.
\section{Conclusion}
\label{sec:conclusion}
Because of their widely tunable properties, 
metallic QDs assembled in regular nanoarrays represent ideal controllable systems for 
bridging the regimes of weak and strong correlations.
The present study of ionization in metallic QD nanorings confirms 
the important impact of electron correlations found in a series of previous 
studies \cite{Baldea:2002,Baldea:2004a,Baldea:2007,Baldea:2008}. 
It demonstrates that the MO-picture of 
ionization in tunable metallic QD-nanorings completely breaks down. 
This breakdown affects \emph{all} MOs.
In ordinary small molecules, it is possible to separate the ionization 
spectra of valence electrons in two distinct regions, related to outer- and inner-valence 
electrons \cite{Cederbaum:1977b,Cederbaum:1980,Cederbaum:1986}. In general, 
the MO-picture holds in the former but breaks down in the latter. 
Such a separation cannot be made in the metallic QD-nanorings investigated here, 
where strong correlations have impact on all of the MOs. 
As a limiting case thereof, even the HOMO--ionization 
is drastically affected by correlations. 
This fact, already pointed out in Ref.\ \onlinecite{Baldea:2002},
contrasts to the case of ordinary molecules. Nevertheless, there exists 
a similarity between QD--nanorings and ordinary molecules: as discussed in Sec.\ 
\ref{sec:10-QDs}, the higher ``occupied'' MOs are less affected by electron correlations 
than the lower ones.
\par
In ordinary molecules, weaker electron correlations manifest themselves in ionization spectra 
as satellite lines of small intensities accompanying the main ionization 
signals \cite{Siegbahn:1967,Siegbahn:1969,Siegbahn:1982,Turner:1970,Rabalais:1977,Berkowitz:1979}.
The main, more intense lines are the result of 
one-hole ($1h$) processes related to the ejection of an electron from an MO.
The satellite, less intense lines are related to excitations accompanying 
the main ionization \cite{Aberg:1970}, 
often consisting of two-hole--one-particle ($2h$--$1p$) processes.
In the case of strong correlations in molecules 
the intensity is distributed over numerous lines with comparable intensity. 
This effect was termed the breakdown of the molecular orbital picture 
of ionization \cite{Cederbaum:1977b,Cederbaum:1980,Cederbaum:1986}.
Main lines cannot be identified in this case, because the properties of 
an ordinary molecule are not tunable, and tracing back to the uncorrelated 
limit is impossible. 
\par
The distinctive feature found in metallic QD-nanorings 
studied here is that, as visible in 
Figs.\ \ref{fig:A1-E1-6-QDs} and \ref{fig:A1-E1-E2-10-QDs},
in the presence of strong correlations the signals originating from 
the main lines of the uncorrelated case progressively lose intensity 
and become dominated by those that trace back to $2h$--$1p$ processes  
in uncorrelated nanorings. 
As concerns the number of lines in the ionization spectra, 
it remains astonishingly scarce. Along with the similar results of 
the recent studies on optical absorption \cite{Baldea:2007,Baldea:2008}, 
this finding gives further support 
to the existence of a hidden quasi--symmetry in the metallic 
strongly correlated QD-nanorings described 
within the extended Hubbard model. As shown here, it remains a 
\emph{quasi}--symmetry even in the case of the restricted Hubbard model ($V=0$). 
\par
With regards to the hidden quasi-symmetry, amply documented here 
and in other recent works \cite{Baldea:2007,Baldea:2008} by detailed numerical results,
but for which a physical explanation is not yet available, we make the following remark.
Exact numerical results played an 
important role in many cases, even including the Hubbard model. The fact 
that the one-dimensional restricted Hubbard model possesses accidental 
degenerqacies was amply discussed in the literature in the past. 
What initiated this issue were the exact numerical results published long time ago 
for benzene-like rings \cite{HeilmanLieb:71}. Although a physical explanation 
could not be given that time, the finding on accidental degeneracies became part 
of wisdom for the Hubbard community. It was only much later that a 
demonstration that these ``accidental'' degeneracies follow from certain nontrivial 
conservation laws could be given \cite{Altshuler:02}. Similarly, we hope 
that our numerical findings will stimulate further investigations 
to unravel the nature of the hidden quasi-symmetry put forward here.
\par
We think that, 
the results of the present theoretical study on ionization
along with those on optical absorption \cite{Baldea:2007,Baldea:2008}
are of interest and hope that they will stimulate 
scientists to fabricate and 
investigate metallic QD-nanorings.
\section*{Acknowledgements}
The authors acknokledge with thanks the financial support provided by the 
Deutsche For\-schungs\-ge\-mein\-schaft (DFG).

\begin{thebibliography}{999}
\bibitem{MO-symmetries}
One should remark here that not all the symmetries of the group D$_{Nh}$ are
  spanned by the MO-symmetries. This is the case, for example, of the
  eigenstates with $A_{2g}$ or $B_{2u}$ of a six-dot nanoring, which cannot
  therefore be studied by photoionization.
\bibitem{total-spin}
The fact that 
$\vert \tilde{\Psi}_{k,\sigma}\rangle  \equiv c_{k,\sigma}\vert \Phi \rangle$ is a state of well 
defined $S_z(=\mp 1/2)$ and $S(=1/2)$ straightforwardly follows from the commutation relations of 
$c_{k,\sigma}$ and the spin operators, and the 
fact that $\hat{S}_z \vert \Phi \rangle = \hat{S}_{+} \vert \Phi \rangle = 0$ 
($\Phi$ is a singlet).
The commutation relation $[\hat{S}_z, c_{k,\downarrow } ] = \frac{1}{2}c_{k,\downarrow}$ 
yields 
$S_z\vert \tilde{\Psi}_{k,\downarrow} \rangle = + \frac{1}{2}\vert \tilde{\Psi}_{k,\downarrow} \rangle $.
Similarly, because $ c_{k,\downarrow}$ and the spin raising operator $\hat{S}_{+}$ commute, 
one immediately gets $S_{+}\vert \tilde{\Psi}_{k,\downarrow} \rangle = 0$ and thence 
$\hat{S}^2 \vert \tilde{\Psi}_{k,\downarrow} \rangle = 
(\hat{S}_z^2 + \hat{S}_{-} \hat{S}_{+} + \hat{S}_{z})\vert \tilde{\Psi}_{k,\downarrow} \rangle = 
\frac{3}{4} \vert \tilde{\Psi}_{k,\downarrow} \rangle$.
\bibitem{avoided-crossings}
The bright diabatic state approximates the mate with significant spectroscopic
  factor of a pair of states involved in an avoided crossing. For instance, in
  Fig.\ \ref{fig:E2-6-QDs}, it describes the state $1$ for $d<1.765$ and the
  state $2$ for $d>1.765$.
\bibitem{scf-poor}
Unless $d$ is very close to unity, using the SCF-ground state $\vert
  \Phi_{SCF}\rangle$ instead of $\vert \Phi \rangle$ is a very poor
  approximation, a fact easily understandable in view of the strong
  correlations.
\end{thebibliography}

\end{document}